\documentclass[fleqn,usenatbib]{mnras}

\usepackage[T1]{fontenc}
\usepackage{ae,aecompl}
\usepackage{afterpage}

\usepackage{graphicx}	% Including figure files
\usepackage{amsmath}	% Advanced maths commands
\usepackage{amssymb}	% Extra maths symbols
\usepackage{nicematrix}

\newcommand{\be}{\begin{equation}}
\newcommand{\ee}{\end{equation}}
\newcommand{\ms}{_{\odot}}

\title[GW-mergers from flyby-perturbed wide binaries]{Detailed properties of gravitational-wave mergers from flyby perturbations of wide binary black holes in the field}

\author[Raveh, Michaely \& Perets]{
Yael Raveh$^1$,\thanks{E-mail: yael.raveh@campus.technion.ac.il}
Erez Michaely$^2$
and Hagai B. Perets$^1$
\\
$^1$ Faculty of Physics, Technion -- Israel Institute of Technology, Haifa, 3200003, Israel\\
$^2$ University of California Los-Angeles 
}

\date{Accepted XXX. Received YYY; in original form ZZZ}

\pubyear{2022}

\begin{document}
\label{firstpage}
\pagerange{\pageref{firstpage}--\pageref{lastpage}}
\maketitle

\begin{abstract}
%No more than 250 words.
Wide black hole binaries (wide-BBHs; $\geqslant 10^3 AU$) in the field can be perturbed by random stellar ﬂybys that excite their eccentricities. Once a wide binary is driven to a sufficiently small pericenter approach, gravitational wave (GW) emission becomes signiﬁcant, and the binary inspirals and merges. In our previous study, using simplified models for wide-BBHs, we found that successive flybys lead to significant merger fractions of wide-BBHs in less than Hubble time, making the flyby perturbation mechanism a relevant contributor to the production rate of GW-sources. However, the exact rates and detailed properties of the resulting GW sources depend on the wide binary progenitors. In this paper we use detailed population synthesis models for the initial wide-BBH population, considering several populations corresponding to different natal-kick models and metallicities, and then follow the wide-BBHs evolution due to flyby perturbations and GW-emission. We show that the cumulative effect of ﬂybys is conductive for the production of GW sources in non-negligible rates of $1-20$ Gpc$^{-3}$ yr$^{-1}$, which are sensitive to the natal kicks model. Such rates are relevant to the observationally inferred rate. Our models, now derived from detailed population of binaries, provide the detailed properties of the produced GW-sources, including mass-functions and delay times. The GW mergers are circularized when enter the aLIGO band; have a preference for high velocity dispersion host galaxies (in particular ellipticals); have a relatively uniform delay-time distribution; and likely have mildly correlated (less than isolated evolution channels and more than dynamical channels) prograde spin-spin and spin-orbits.
%We expect these mergers to show no signiﬁcant spin–orbit alignment, and uniform delay-time distribution.
%We first generate different populations of BBHs, considering several models for BH natal-kicks.  We then implement a code that simulate their long-term evolution of the wid.

\end{abstract}


\section{Introduction}
The first direct detection of gravitational-wave (GW) sources, and the tens of identified sources found since, opened a new era in the study of GW sources and compact objects. There are 90 reported LIGO/Virgo/KAGRA (LVK) gravitational-wave detections \citep{2021arXiv211103606T} and the observational data will be rapidly increasing with the next observing runs of LVK collaboration, enabling novel ways to test the astrophysical theories explaining the origin of these sources. The currently inferred stellar binary black hole (BBH) merger rate from these observations in the local Universe is $17.3-45 $Gpc$^{-3}$ yr$^{-1}$; while the merger rates of neutron star (NS) binary is $13-1900$ Gpc$^{-3}$ yr$^{-1}$, and of NSBH is $7.4-320$ Gpc$^{-3}$ yr$^{-1}$ \citep{2021arXiv211103634T}.

Various scenarios were proposed for the origins of GW sources from mergers of BBHs, including isolated evolution of massive binaries in the field \cite[e.g.,][]{2016ApJ...819..108B,2015ApJ...806..263D}, dynamical channels involving close encounters in dense clusters \cite[e.g.,][]{2016PhRvD..93h4029R,Stephan2016,2018PhRvL.121p1103F,2018MNRAS.474.5672L,Hoang2020}, secular dynamical evolution of triple or higher multiplicity systems \cite[e.g.,][]{2017ApJ...841...77A,2012ApJ...757...27A,2018PhRvD..97j3014S}, and gas-assisted mergers of compact binaries \cite[e.g.,][]{2018PhRvL.120z1101T,2012MNRAS.425..460M,2022arXiv220301330R}. 
Large uncertainties exist in all of these models, and the initial predicted GW-production rates range over a few orders of magnitudes below and above the currently inferred rates. The new measurements do constrain the parameter ranges of all suggested models, but can not, currently, exclude at least some parameter ranges of the models. In particular, it is possible that more than one channel contribute to the rates, although, unless fine-tuned, it is likely that a single channel would dominate over the others. 

%such as Lidov-Kozai resonances in hierarchical triples \cite[e.g.,][]{2017ApJ...841...77A,2012ApJ...757...27A,2018PhRvD..97j3014S}, dynamical capture in open/globular/nuclear clusters , chemically homogeneous evolution in compact stellar binaries \cite[e.g.,][]{2016MNRAS.458.2634M}, and mergers of primordial black holes \cite[e.g.,][]{2016ApJ...819..108B,2015ApJ...806..263D}. 

In a novel scenario first suggested by us in a series of papers \citep{2016MNRAS.458.4188M,2019ApJ...887L..36M,mic+20,michaely2021a,michaely2021}, we describe the collisional interaction of wide systems in the field of the host galaxy as source of binary exotica. Specifically, for GW sources  \citep{2019ApJ...887L..36M}, wide (SMA$>$1000 AU) BBHs in the ﬁeld are perturbed by random ﬂyby interactions of ﬁeld stars in their host galaxy. Like the isolated evolution models, the progenitors are located in the field and do not require high stellar density environments such as clusters and galactic nuclei. However, unlike the isolated and secular evolution models, these models are catalyzed by collisional evolution through encounters with stars, more similar to the dynamical channels in cluster environments. We also note that secular evolution could play a role in the evolution of binaries and triples in the field, where galactic tides play the role of an external perturber, in somewhat similar way as the perturbation of a third star on the inner binary through van Zeipel-Lidov-Kozai oscillations \cite[see][and references therein]{gri+22}. Here we do not consider the effects of galactic tides and focus only on flyby encounters.

As the paper suggests, due to interaction with stellar perturbers, a fraction of wide-orbit BBHs can be driven into sufficiently small pericenter distances in order to merge via GW emission within a Hubble time and serve as GW-sources detectable by VLK \citep{2019ApJ...887L..36M}. We derived an analytic model and verified it with numerical calculation to compute the merger rate to be $\sim1\times f_{wide} Gpc^{-3} yr^{-1}$ ($f_{wide}$ is the fraction of wide eccentric BBHs), which is a relevant contribution to the observationally inferred rate. This study considered only spiral galaxies, and not included the contribution from wide triples. We have later shown that both of these could potentially increase the rates up to a few tens ${\rm Gpc}^{-3} yr^{-1}$, i.e. comparable to the currently inferred GW rate. 
%Three commonly acknowledged evolutionary channels were proposed in the context of GW mergers; merger in dense environments, the isolated evolution of initially massive close binary stars (post-CE binaries), and mergers induced by secular evolution of triple systems either in the ﬁeld or in dense environments.
%\cite{2019ApJ...887L..36M} presented a fourth evolutionary channel, in which they focused on wide (SMA$>$1000 AU) black hole binaries (BBHs) in the ﬁeld perturbed by random ﬂyby interactions of ﬁeld stars in their host galaxy. They found that due to interaction with stellar perturbers, a fraction of wide-orbit BBHs can be driven into sufficiently close pericenter distances to merge via GW emission within Hubble time and are thus detectable by aLIGO/VIRGO.

Nevertheless, the merger rate we calculate \citep{2019ApJ...887L..36M}, strongly depends on the natal kicks (NKs) given to BHs, which are poorly constrained \citep{2012MNRAS.425.2799R}. In particular, it is still unknown whether a BH receives a momentum kick at birth like a NS, or forms without any natal-kick following a failed supernova or a large amount of fallback \cite[e.g.,][]{2008ApJS..174..223B,2016ApJ...818..124E}. In \cite{2019ApJ...887L..36M} we assumed zero kick velocities (as done by other potentially successful scenarios).  As noted there, a high production rate of GW sources through the wide-binaries channel might be compromised if non-zero NKs are considered, since these can unbind initially wide binaries and thereby lower their fraction and the overall production rate of GW sources. Therefore, in the previous studies an initial population of wide binaries/triples were assumed, and low-kicks were required in order for such systems to survive. 

Here we relax the assumption of very low NKs, and show that the wide-binary channel does not require very low NKs, and in fact, NKs could in fact serve to catalyze the formation of wide binaries from shorter period binaries. In particular,  a shorter period binary ($<1000$ AU) receiving a kick, with comparable velocity to its orbital velocity could become unbound, but could also just be excited to very large semi-major axis and eccentricity as to become a wide ($>1000$ AU) binary. In other words, the wide-binary model might not require very small NKs in order to produce the wide binary progenitors, but rather the formation of wide binaries could be assisted by non-negligible NKs. 

Here we study this possibility in detail by studying self-consistent formation models of wide binaries arising from shorter period binaries and their detailed properties. We then use the wide-binaries as to study the resulting GW-sources formed through flyby perturbation in the field of the wide-binary populations studied. 

We calculate the merger probability for BBHs that are formed through such scenario by first generating different populations of BBHs, considering several typical NK models discussed in the literature. This is done by modelling isolated binary evolution with the population synthesis code COMPAS \citep{2017NatCo...814906S,2018MNRAS.481.4009V,2019MNRAS.490.5228B}. We then estimate the fractions of GW-mergers among the outputted wide BBHs by following their long-term evolution due to random flyby perturbations for $10$ Gyr, and identifying the cases where the binaries merge in less than a Hubble time. Given the random nature of the flyby perturbations, the probability for a merger is obtained through a Monte-carlo statistical analysis. Namely, we run a large number of realizations of the random flyby perturbation evolution for each of the initially wide and eccentric binary (non eccentric binaries are unlikely to be sufficiently perturbed as to merge, and are ignored to save computational expenses) outputted from our isolated evolution population synthesis model. We then calculated the fraction of realizations leading to mergers, which provide us with the merger probability. The contributed sum of probability weighted mergers then provides us with the estimated production rates of GW sources and the properties. 
%using the parametric binary population synthesis code named COMPAS. %the same binary set up and stellar environment, (speciﬁcally for a BBH with = pertubed by stars of typical mass = and velocity dispersion of =. 

The study is structured as follows. In Sec.~\ref{sec:method} we explain the method used to generate a Monte-Carlo population of isolated field BBHs. In Sec.~\ref{sec:flyby} we present the numerical scheme used to simulate dynamical interactions of random flybys with all outputted wide eccentric BBHs.
In Sec.~\ref{sec:results} we present our main results and make detailed predictions for observing runs of the future GW detector network%, distinguishing three regions of the BBH parameterspace
. We then discuss and conclude our results in Sec.~\ref{sec:sum} %where we compare our work to the current literature and demonstrate the importance of detailed binary evolution calculations
. Finally, the conclusions of our work are given in Sec.~\ref{sec:con}.

\section{Methods} 
We focus on wide BBHs with semimajor axis (SMA) $a > 10^3$ AU. Each binary resides in the field of a host galaxy and is therefore affected by short duration dynamical interactions with field stars. %These perturbations can change the binary SMA, $a$ and the binary eccentricity, $e$. It has been shown (Lightman & Shapiro 1977; Merritt 2013) that angular momentum change is more significant than energy change for this process, hence we focus on the eccentricity change. 
As discussed in \cite{2019ApJ...887L..36M}, if such interactions produce a sufficiently small pericenter passage, then the system can merge via GW emission within Hubble time. 

\subsection{COMPAS model assumptions} \label{sec:method} 
\begin{figure*}
\includegraphics[scale=0.58,clip]{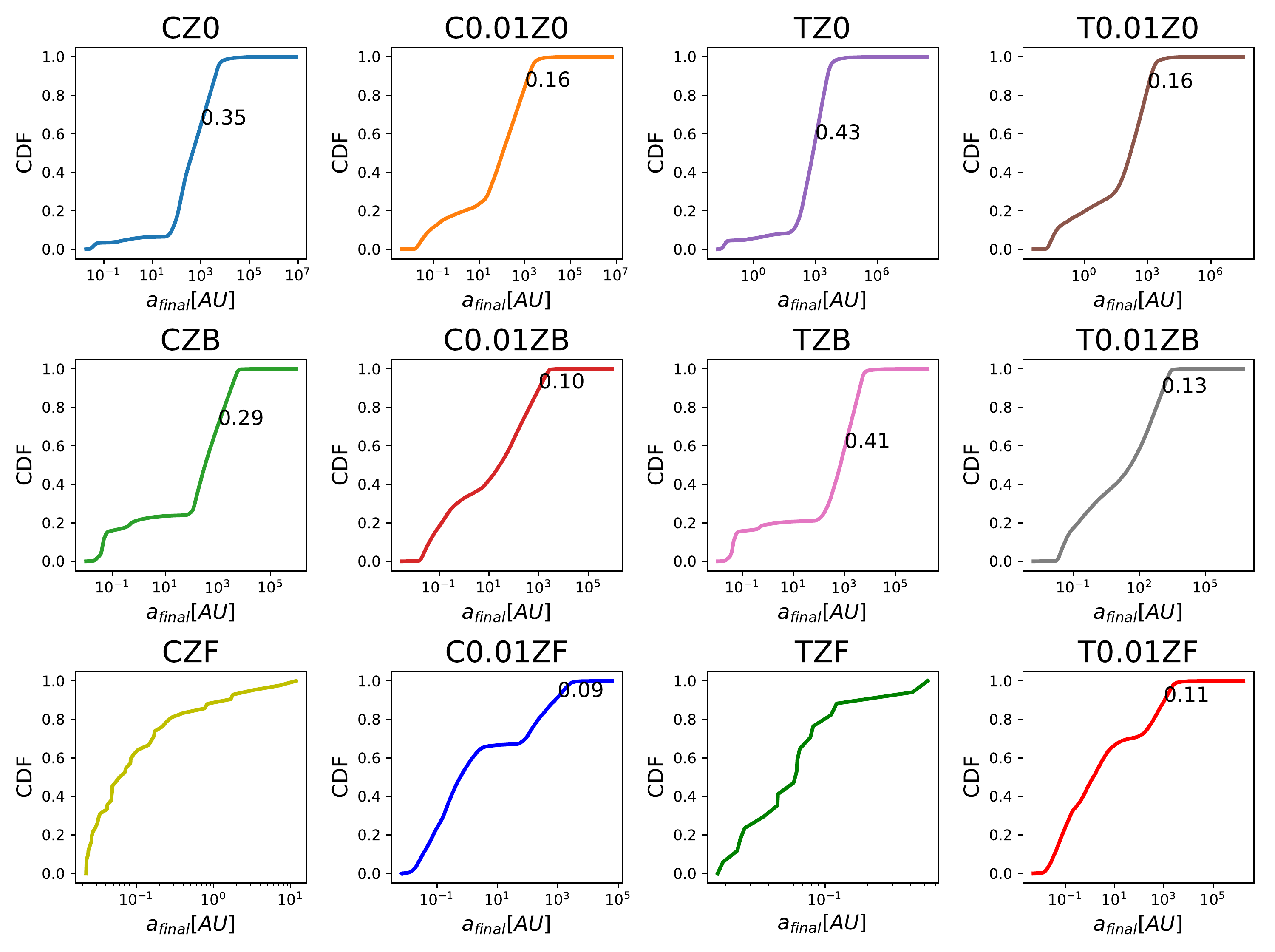}
\caption{Separation CDF for the BBH populations (acronyms are explained in Table~\ref{tab:grid}), considering different natal-kick models, once they are formed in COMPAS.
} \label{fig:sma}
\end{figure*}
\begin{figure*}
\includegraphics[scale=0.58,clip]{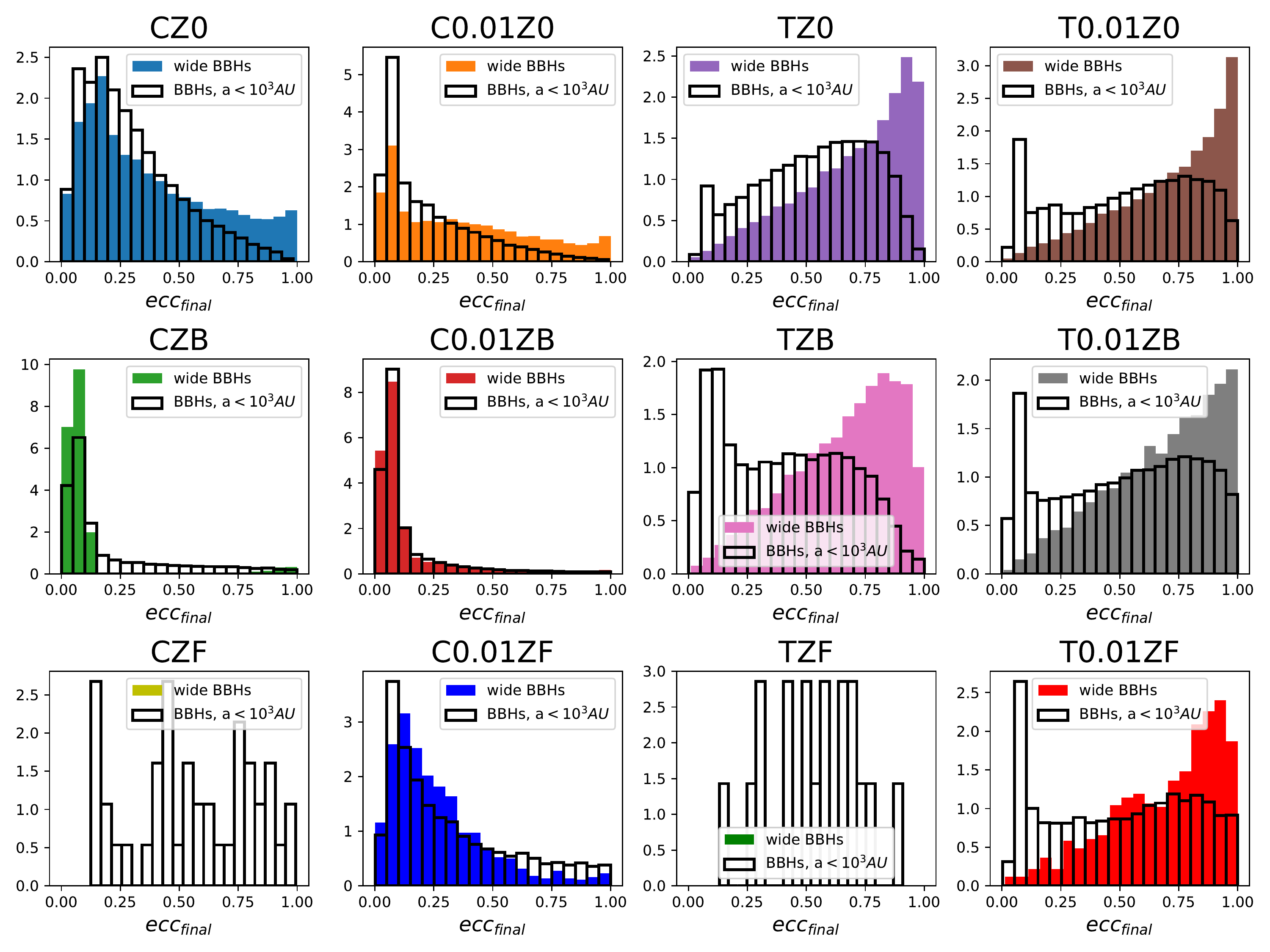}
\caption{Eccentricity profiles for the BBH populations (acronyms are explained in Table~\ref{tab:grid}), considering different natal-kick models, once they are formed in COMPAS.
} \label{fig:ecc}
\end{figure*}
\begin{figure*}
\includegraphics[scale=0.58,clip]{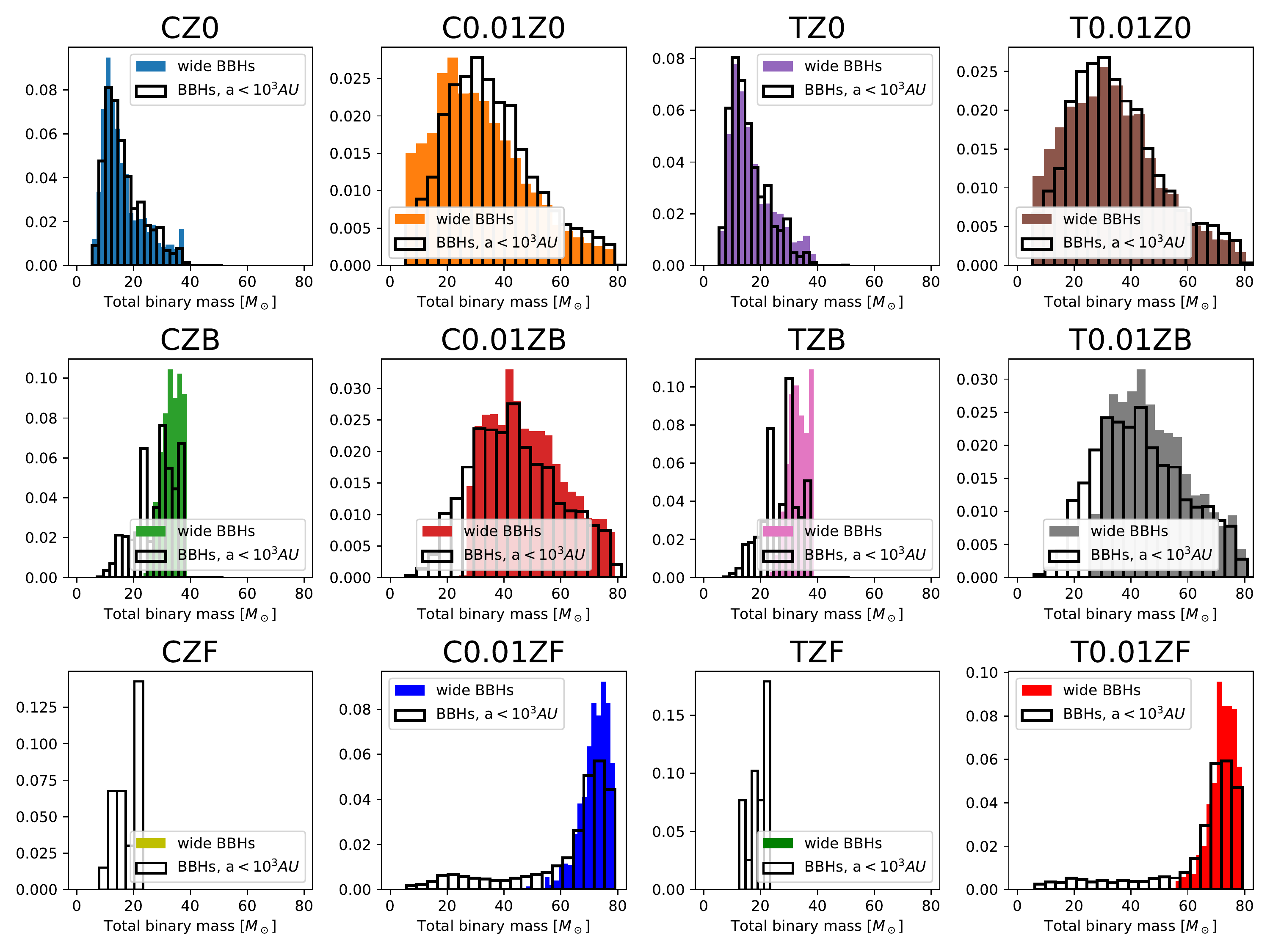}
\caption{Mass profiles for the BBH populations (acronyms are explained in Table~\ref{tab:grid}), considering different natal-kick models, once they are formed in COMPAS.
} \label{fig:mass}
\end{figure*}
\begin{figure*}
\includegraphics[scale=0.58,clip]{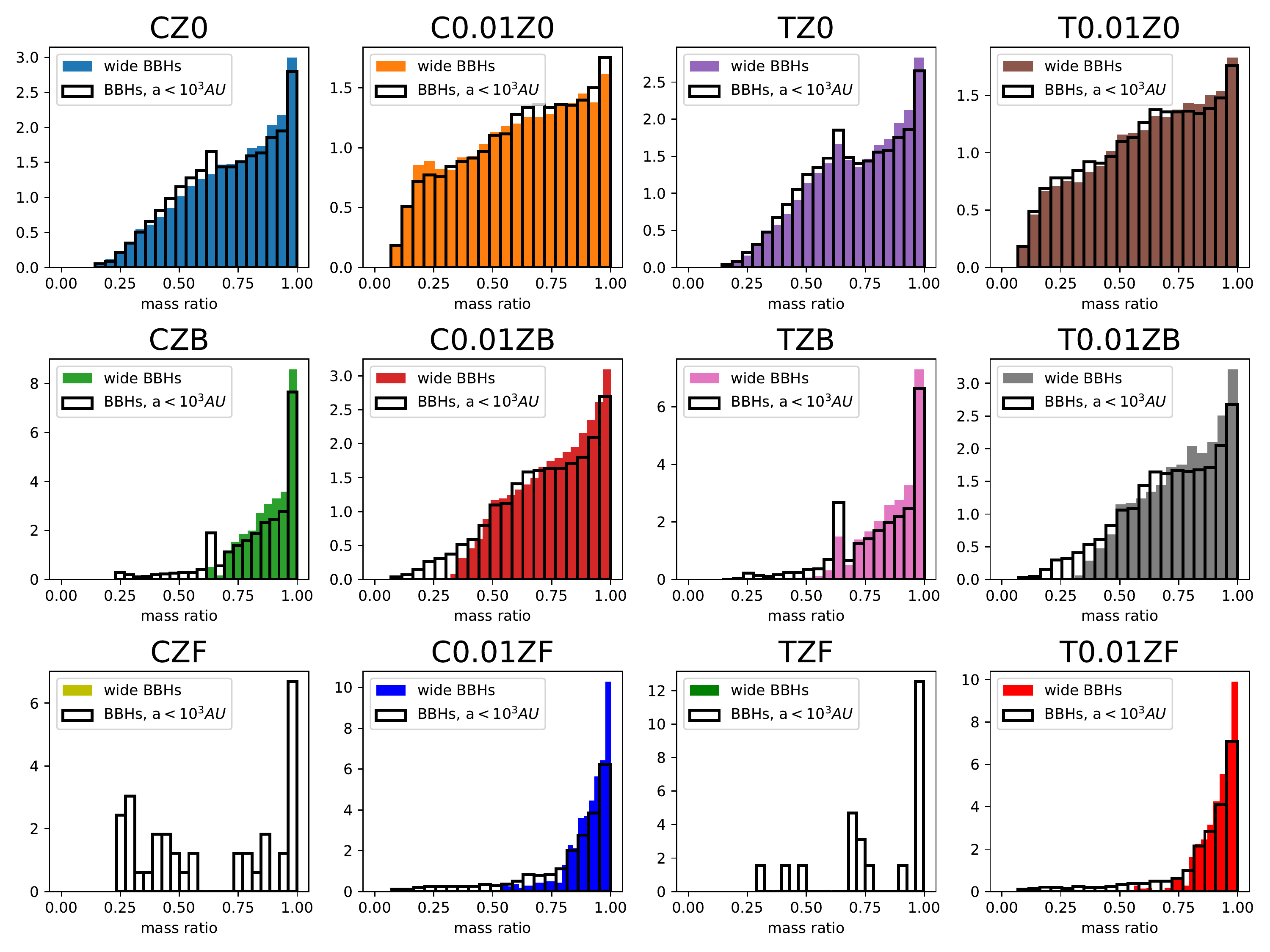}
\caption{Mass-ratio profiles for the BBH populations (acronyms are explained in Table~\ref{tab:grid}), considering different natal-kick models, once they are formed in COMPAS.
} \label{fig:q}
\end{figure*}
We consider the evolution of an ensemble of wide BBHs with initial separations $a>10^3$~AU.
In order to obtain the statistics of large populations, we used the publicly available open code COMPAS \citep{2017NatCo...814906S,2018MNRAS.481.4009V,2019MNRAS.490.5228B}. %The parameters used are presented in Table~\ref{tab:grid}. 
The initial conditions for the binary population were constructed as in \cite{2019MNRAS.490.3740N}, with the following main physical assumptions;
since we are interested in binaries with a primary star that ends up forming a BH, we drew $n_b = (3-12)\times 10^6$ binaries with primary mass $5 M\ms < m_1 < 150 M\ms$ from a Kroupa initial mass function \cite{2001MNRAS.322..231K}. The mass distribution of the less massive secondary star is given by $m_2 = m_1 q$ where $q$ is the initial mass ratio 

($0 < q < 1$) drawn from a flat distribution \citep{2012Sci...337..444S}. We assume that, at formation, binaries are distributed uniformly in log-orbital separation restricted to $0.1 < a/AU < 1000$ \citep{1983ARA&A..21..343A}. Few test runs were performed for a log-orbital separation restricted to $0.1 < a/AU < 50000$. The eccentricities of the binaries were chosen either as zero or from a thermal distribution, i.e. $f(e) = 2e$ (see Table~\ref{tab:grid}). We assume that all distributions are both independent of each other as well as independent of metallicity, where in this study we consider only solar metallicity, namely $Z_\odot = 0.0142$ \citep{2009ARA&A..47..481A}, or $Z = 0.01Z_\odot$.

COMPAS evolves stars according to the stellar models of \cite{8135443} and uses analytical fits of these models to rapidly evolve binaries \citep{2000MNRAS.315..543H,2002MNRAS.329..897H}, such that eventually, only a small fraction of interacting massive binaries form BBHs. This requires stars to avoid merger during mass transfer; to have sufficient mass to form BHs; the binary must remain bound through SNe; and after the formation of a BBH, the binary must be tight enough to merge within the age of the Universe and create a detectable GW event. Concerning the latter, we argue that wide BBHs in the ﬁeld, which are perturbed by random ﬂyby interactions of ﬁeld stars in their host galaxy, can actually be driven into sufficiently close pericenter distances to merge via GW emission within Hubble time. These flyby interactions are not considered in the COMPAS model, and are studied in the next step, as we discuss below in section \ref{sec:flyby}. 
According to \cite{2016arXiv161000593M}, the rates for such scenario strongly depend on the natal-kicks given to BHs, which are poorly constrained. Hence we calculate the rates considering three NK models as detailed below. %In one scenario, we apply a NS-like prescription for the kicks. In another scenario we assume the fallback kick prescription.

\subsection{Natal kicks}
Supernovae explosions can be asymmetric and, as a result, impart a kick on the formed remnant. In order to produce the natal kick velocity vector we randomized the kick velocities chosen from a Maxwellian distribution with a velocity dispersion of $\sigma = 265$ kms$^{-1}$ \citep{2005MNRAS.360..974H}, and isotropic unit vector around the NS. The fraction $f_b$ of mass that falls back onto the newly born compact object is prescribed by \cite{2012ApJ...749...91F}. Then, the natal kick is proportionally reduced based on the fallback fraction according to
\be v_{kick} = (1-f_b)v_{kick,drawn}. \ee

The merger rate we calculate below strongly depends on the natal-kicks given to BHs, which are poorly constrained \citep{2016arXiv161000593M}. Previous models that were able to produce rates comparable to the rate inferred from observations, have typically assumed zero kick velocities (for all BHs, or at least for all BHs more massive than $10 M\ms$), while models assuming higher natal kicks produced signiﬁcantly lower rates. 

We conclude that adapting similar no-kick assumptions for BHs (as done by other potentially successful scenarios), i.e. $f_b=1$, suggests the wide-binary channel explored here can give rise to a high production rate of GW sources from BBH mergers of perturbed ultra-wide binaries. Nonetheless, we performed our calculations considering two more natal kicks models; one in which $f_b=0$, referenced therein as the full NK model, and another - the fallback model - where $f_b$ is mass dependant and is determined according to \cite{2012ApJ...749...91F}.

\subsection{Mergers of BBHs catalyzed by flybys in the field} \label{sec:flyby} 
To simulate flyby perturbations to the binaries outputted from COMPAS, we adopt a numerical scheme firstly suggested and used by  \cite{2019ApJ...887L..36M}. Here we highlight the main physical assumptions.

We have simulated the evolution of the wide eccentric BBHs ($a>1000$ AU ; $e>0.98$; lower eccentricity binaries are unlikely to evolve to sufficiently high eccentricities as to merge, and were assumed not to produce any mergers, as to save computation time) that were outputted from COMPAS for $10$ Gyr by considering both the evolution of the binary between encounters, and in particular the effects of GW emission, as well as the change of the binary orbital elements due to the impulsive flyby encounters with field stars. 

For each ensemble of NK model, we calculated the fraction that merged out of the initial binaries, as well as the eccentricity at the %LISA and 
LIGO band. 

The dynamical encounters are modeled through the impulse approximation\footnote{Farther out, flybys can also perturb the system and further excite the system, but at very large separations the interaction becomes adiabatic and the effects become small. In addition, we neglect the intermediate regime in which the perturbation time and the orbital times are comparable, which are likely to somewhat enhance the perturbation rates explored here.}, namely $t_{int}\equiv b/v_{enc}\ll P$, where $b$ is the closest approach to the binary, $v_{enc}$ is the velocity of the perturbing mass as measured from the binary center of mass, and $P$ is the orbital period. %There are actually three more relevant timescales for the impulsive treatment: the binary orbital period $P$; the merger time from a specific binary configuration via GW emission $t_{merger}$; and the time between two consecutive encounters of the system and a flyby perturber, $t_{enc}= 1/f = (n_*\sigma v_{enc})^{-1}$, where $n_*$ is the stellar number density, and $\sigma$ is the geometric cross-section of the binary and the flyby.

To calculate the average time between encounters we use the rate $f = n_0\sigma \langle v_{enc}\rangle$, where in our fiducial model $n_0$ is the stellar number density, taken to be the solar neighborhood value of $n_0 = 0.1$ pc$^{-3}$, $\sigma$ is the interaction cross-section, and we set $\langle v_{enc}\rangle = 50$ kms$^{-1}$ similar to the velocity dispersion in the solar neighborhood. Later we consider other environments; in Section \ref{sec:dense} we present a test case of denser fields within spiral galaxies with $\langle v_{enc}\rangle = 50$ kms$^{-1}$ and $n_0 = 1.5$ pc$^{-3}$, while Section \ref{sec:ell} deals with test cases in elliptical galaxies with $\langle v_{enc}\rangle = 160$ kms$^{-1}$ and $n_0 = 0.1$ pc$^{-3}$.

Since we focus on the impulsive regime and consider the values mentioned above, the largest closest approach distance $b$ for which an encounter can be considered as impulsive is $b_{max} = 5\times 10^4$ AU. The average time between such impulsive encounters is given by $t_{enc} = f^{-1} \approx 1 Myr$. %Therefore we randomly sample the time between encounters from an exponential distribution with a mean $f$ (due to the Poisson distribution of encounter times).

For each binary extracted from COMPAS, we recorded the SMA $a$ and eccentricity $e$. The orbit of each such wide binary resulting from COMPAS is then taken as the initial orbit of the wide eccentric BBH in our flyby model. At each step we first find the next encounter time $t_{enc}$, and evolve the binary up to that point through the well known equations of motion given by \cite{PhysRev.136.B1224}, i.e. accounting for the orbital evolution due to GW emission. If the binary did not merge through GW emission by the time of the next encounter, we simulate the next impulsive interaction with a perturber with velocity of $v_{enc}$ drawn from a Maxwellian distribution with velocity dispersion $\langle v_{enc}\rangle$ and a mass of $m_p = 0.6M\ms$, typical for stars in the field. After changing the binary orbital parameters due to the encounter, we continue to evolve the binary until the next encounter and so on, up to the point when the binary merges, is disrupted, or the maximal simulation time of $10$ Gyrs is reached.

To calculate the post-interaction orbital elements of a perturbed binary, we first randomize the perturber trajectory; we randomly sample the closest approach point $\mathbf{b}=b\hat{b}$ of the perturber from an isotropic distribution. Here $b$ is uniformly distributed in the range $(0,b_{max})$. %by randomizing the two spherical coordinate angles, uniformly distributed $cos\alpha \in (-1,1)$ and uniformly distributed $\beta\in (0,2\pi)$, which determine a plane perpendicular to the closest approach vector $b$. The perturber trajectory can have any direction within this plane, hence we randomize an additional angle, uniformly distributed in the range $\gamma\in (0,2\pi)$ as to choose an arbitrary axis in the plane.
%Next we randomize $b$ by setting the distribution function to be $f(b)\propto b_{max}$.

We calculate the post-interaction binary orbital elements using the impulse approximation. In the impulse approximation %we neglect any motion of the binary during the passage of perturber. In this approximation 
we can calculate the velocity change of each of the components of the binary, $\mathbf{\delta v_1}, \mathbf{\delta v_2}$, as described in \cite{2008AJ....136.2552C,2019ApJ...887L..36M}. %For $m_1$, the velocity vector change is given by
%where $\mathbf{r_p}$ is the position of the flyby perturber set to be at the closest approach at $t = 0$. Since it is the impulse regime, we approximate the trajectory of the flyby during the interaction as a straight line thus $\mathbf{r_p}=\mathbf{b}+\mathbf{v_{enc}}t$. The velocity change for $m_2$ is then
Given the relative velocity, $\mathbf{v_r}$ (as a function of the SMA, eccentricity and the mean anomaly), and the relative velocity change, $\mathbf{\delta v_r} = \mathbf{\delta v_1} - \mathbf{\delta v_2}$, we can calculate the post-interaction eccentricity vector
\be \mathbf{e}\equiv \frac{(\mathbf{v}+\mathbf{\delta v_r})\times (\mathbf{r}\times (\mathbf{v}+\mathbf{\delta v_r}))}{Gm_b}-\frac{\mathbf{r}}{r} \ee
where $r$ is the separation and $m_b$ is the binary mass. i.e.,
\be e_{post}=\vert \mathbf{e}\vert. \ee

The post-interaction SMA is calculated via the change in orbital energy. The energy change is due to the velocity kick imparted by the perturber. In the impulse approximation we model the interaction via a velocity change. The specific orbital energy is
\be \epsilon = \dfrac{v^2}{2}-\dfrac{Gm_b}{r}=-\dfrac{Gm_b}{2a}. \ee
The specific energy change is given by
\be \delta\epsilon = \frac{\mathbf{\delta v_r}^2+2\mathbf{v}\cdot\mathbf{\delta v_r}}{2}, \ee
which translates to change in the SMA of
\be \delta a = -a\dfrac{\delta\epsilon}{\epsilon}, \ee
to give us the final post-interaction SMA 
\be a_{post} = a+\delta a. \ee

For each of the modeled initial conditions, we calculate the fraction of merged systems after $t = 10$ Gyr since the BBH was initialized. More specifically, we record the number of merged systems $n(a,e)$ to find the fraction of merged systems out of the total number of modeled systems $f_m=n(a,e)/N(a,e)$, where $N(a,e)$ is $\mathcal{O}(10^3-10^4)$. In Table~\ref{tab:grid} we show averaged per NK model merger ratios of all modeled wide BBHs.

\subsection{Rate estimates of wide BBH mergers} \label{sec:rate}
In order to estimate the volumetric rate of GW mergers from the mechanism in question, we ﬁrst calculate the number of systems merging in a Milky Way (MW)-like galaxy per unit time, as done in \cite{2019ApJ...887L..36M} and references therein. %In order to do that we need to integrate over all SMA in a given stellar density and over all stellar densities in the galaxy we model. We follow 
Following a similar calculation, we consider $dN = 2\pi n_*(r) h r dr$ to be the number of stars in a region $dr$ (and scale-height $h = 1$~kpc), located at distance $r$ from the center of the Galaxy. The Galactic stellar density in the Galactic disk is modeled as $n_*^{spiral}(r) = n_0e^{(-(r-r_\odot)R_l)}$, where $n_0 = 0.1$~pc$^{-3}$ is the stellar density near our Sun, $R_l = 2.6$~kpc \citep{2008ApJ...673..864J} is the galactic length scale and $r_\odot = 8$~kpc is the distance of the Sun from the galactic center.

Integrating over the stellar densities throughout the disk of the Galaxy ($r=0.5-15$~kpc), and using our COMPAS results, we can obtain the total yield of wide BBHs mergers in the Galaxy through the process of random flyby interactions. We account for the fraction of wide BBH systems from the entire population of stars in the Galaxy, $f_{BBH}$, by calculating the ratio between the number of wide eccentric BBHs $n_{wide}$ and the overall number of binaries we simulated in COMPAS $n_b$. We additionally re-normalize the population synthesis simulation to the total stellar mass of the underlying stellar population. The normalization  accounts for the fact that the binary population we modeled amounts for only a fraction of the underlying stellar population; i.e. it consists only of binaries of primaries with restricted (high) mass that can serve as BH progenitors (as to save  computational costs). Since we used the \cite{2001MNRAS.322..231K} IMF with $m_{min}=0.01 M\ms$ and $m_{max} = 200 M\ms$ as the minimum and the maximum stellar mass, and the observed binary fraction $f_{bin} = 0.7$ \citep{2012Sci...337..444S}, and since we carried out the simulation for primary masses in the range between $5 M\ms < m_1 < 150 M\ms$ \citep{2005Natur.434..192F}, we obtain $2n_b/n_{total} \sim 0.0072$.

To calculate the overall rate per NK model, we sum over all merger fractions $f_m$ and normalize by $n_{wide}$, i.e. the number of wide eccentric BBHs outputted for the NK model. All in all, the number of merging wide and eccentric BBHs ($a>1000$ AU ; $e>0.98$) per Myr from this channel for a MW-like Galaxy is 
\be R = \int_{0.5}^{15 kpc}\sum_{f_m} \frac{2\pi n_*(r)hr}{10^4 Myr}\left(\frac{2n_b}{n_{total}}\right)f_{BBH}\left(\frac{f_m}{n_{wide}}\right) dr \label{eq:rate} \ee 
Following \cite{2016ApJ...819..108B} we calculate the merger rate $\Gamma^{spiral}$ per Gpc$^{-3}$ per yr by using the following estimate $\Gamma^{sp} = 10^3R\times\rho_{gal}^{spiral}$, where $\rho_{gal}^{spiral}$ is local density of MW-like galaxies with the value of $\rho_{gal}^{spiral} = 0.0116$ Mpc$^{-3}$ \cite[e.g.,][]{2008ApJ...675.1459K} and $R$ is given in units of Myr$^{-1}$.

%\color{red}
%\subsubsection{The finite 
%lifetime of wide binaries due to flybys} \label{sec:rmin}
%HBP: Why do we have this section? the numerical calculation of the flybys already includes their ionization
%Binaries may be ``ionized'' and destroyed, i.e. disrupted, by random ﬂybys in the field. Such an ionization process, which is more frequent in dense stellar environments such as the Galactic center, decreases the available number of wide binaries. To account for the ionization process we consider the ﬁnite lifetime of wide binaries due to ﬂybys using the approximate relation given by \cite{1985ApJ...290...15B} for $t_{1/2}$, the half-lifetime of a wide binary evolving through encounters
%\be t_{1/2}=0.00233\frac{v_{enc}}{Gm_pn_*a}. \ee
%We set $t_{1/2}=10$~Gyr, and find $r=r_{min}$ from which onward we can neglect the aforementioned ionization process such that our numerical scheme and rate estimates stay valid.  
%\color{black}

\subsubsection{Test case of wide binaries in denser fields} \label{sec:dense}
As mentioned above, in the flyby simulation we calculate the average time between encounters as the inverse of the rate $f\propto n_0$, where $n_0 = 0.1$~pc$^{-3}$ is a \emph{constant} stellar number density. This is done for simplicity reasons and assuming the outputted merger probabilities $f_m$ merely weakly depend on the stellar density in the region where the binaries are at in a galaxy. More specifically, the strongest dependence of similar mass wide binaries due to random encounters with flybys in the field according to \cite{2019ApJ...887L..36M}, does not exceed $\propto n_*^{1/2}$. %We therefore obtain merely rough estimates of the merger probabilities $f_m$ of our systems, since $f_m$ depend on the stellar density in the region where the binaries are at. we then use the outputted merger probabilities of all systems $f_m$ as input to Eq.~(\ref{eq:rate}) and calculate the number of merging wide eccentric BBHs per megayear for a MW-like GalaxyThe merger probabilities of wide eccentric BBHs due to random flyby encounters $f_m$ depend however

 In order to consider other environments, we also considered a model with a higher stellar number density density, as expected in the inner parts of spiral galaxies ($n_{dense}\sim1.5$ pc$^{-1}$), and a model with a higher velocity dispersion (160 km s$^{-1}$), as expected in elliptical galaxies (discussed in further detail below). Given the computational cost, these additional environments were only fully simulated for one NK model (CZB), while the ratio between the rates in these different environments in the CZB  models was then used to extrapolate the rates in high density and high velocity dispersion environments in the other models, assuming similar ratios as found in the CZB model.  

We than implemented the new merger probabilities in Eq.~(\ref{eq:rate}) and obtained the overall rate  shown in Table~\ref{tab:grid}.
%We find that increasing the stellar density around the wide binaries by a factor of $\sim 15$ lead to an increased merger probability of $\sim 2.2$. 

\subsubsection{Test case of wide binaries in elliptical galaxies} \label{sec:ell}
Since the different properties of spiral and elliptical galaxies generally affect merger rates (through different stellar number density and velocity dispersion in different types of galaxies), we calculate the merger rate both for typical MW-like spiral galaxies, as done in Section~\ref{sec:rate}, and for elliptical galaxies.

For elliptical galaxies we take the density profile from \cite{1990ApJ...356..359H} and translate it to stellar density given an average stellar mass of $m_{av} = 0.6 M\ms$,
\be n_*^{elliptical}(r)=\frac{M_{gal}}{2\pi m_{av}r}\frac{r_*}{(r+r_*)^3} \label{eq:den} \ee
where $r_* = 1$ kpc is the scale length of the galaxy and $M_{gal} = 10^{11} M\ms$ is the total \emph{stellar} mass of the galaxy. The velocity dispersion for a typical elliptical galaxy we consider is $\sigma = 160$~km s$^{-1}$; that translates to much faster perturbers to wide binaries within elliptical galaxies. Relevant adjustments were made to the flyby simulation to obtain merger ratios $f_m$ of wide eccentric BBHs.

We calculate the merger rate of wide eccentric BBHs from elliptical galaxies in a similar way to the one described in Sections \ref{sec:flyby} and \ref{sec:rate} with $v_{enc} = 160$~km s$^{-1}$ and the stellar density in Eq.~(\ref{eq:den}).

Next, we input the number density of elliptical galaxies in the local Universe $\rho_{gal}^{elliptical} \cong 0.1$ Mpc$^{-3}$ \citep{2014ApJ...784...71S} and get $\Gamma^{el} = 10^3R\times\rho_{gal}^{elliptical}$, where $R$ is given in units of Myr$^{-1}$.

\begin{table*}
    \centering
    \begin{NiceTabular}{|l|c|c|c|c|c|c|c|c|c|}[hvlines]
        \Block{2-1}{\textbf{Acronyms}} & \Block{2-1}{} \textbf{Natal kick} & \Block{2-1}{} \textbf{e$_{init}$} & \Block{2-1}{\textbf{Metallicity}} & \Block{2-1}{\textbf{f$_{wide}$}} & \Block{2-1}{\textbf{f$_{BBH}$}} & \Block{2-1}{} \textbf{merger ratio} & \Block{2-1}{} \textbf{$\Gamma^{sp}_{0.1 pc^{-3}}$} & \Block{2-1}{} \textbf{$\Gamma^{sp}_{1.5 pc^{-3}}$} & \Block{2-1}{} \textbf{$\Gamma^{el}_{0.1 pc^{-3}}$ } 
        %& \Block{2-1}{} \textbf{$\Gamma^{sp,iso}_{0.1 pc^{-3}}$}
        \\
        & \textbf{model} & \textbf{DF} & & & 
        & \textbf{(average $f_m$)} & \textbf{\tiny [Gpc$^{-3}$ yr$^{-1}$]} & \textbf{\tiny [Gpc$^{-3}$ yr$^{-1}$]} & \textbf{\tiny [Gpc$^{-3}$ yr$^{-1}$]} %& \textbf{\tiny [Gpc$^{-3}$ yr$^{-1}$]} 
        \\
        TZF & \Block{4-1}{Full} & \Block{2-1}{Thermal} & Z$_{sol}$ & 0.0 & 0.0 & no wide BBH & - & - & - %& 4 
        \\ 
        T0.01ZF & & & 0.01 Z$_{sol}$ & 0.09 & 4.66e-6 & no wide BBH$^a$ & - & - & -% & 708
        \\ 
        CZF & & \Block{2-1}{Circular} & Z$_{sol}$ & 0.0 & 0.0 & no wide BBH & - & - & - %& 8 
        \\ 
        C0.01ZF & & & 0.01 Z$_{sol}$ & 0.11 & 1.66e-6 & no wide BBH$^a$ & - & - & - %& 1050 
        \\
        TZB & \Block{4-1}{Fallback} & \Block{2-1}{Thermal} & Z$_{sol}$ & 0.41 & 1.71e-5 & 0.002 & 0.029 & 0.064 & 1.01 %& 376 
        \\
        T0.01ZB & & & 0.01 Z$_{sol}$ & 0.13 & 1.05e-4 & 0.003 & 0.246 & 0.545 & 8.63 %& 3439
        \\ 
        CZB & & \Block{2-1}{Circular} & Z$_{sol}$ & 0.29 & 8.41e-6 & 0.0043 & 0.028 & 0.062 & 0.982 %& 557 
        \\ 
        C0.01ZB & & & 0.01 Z$_{sol}$ & 0.10 & 1.36e-5 & 0.0046 & 0.049 & 1.1$^b$ & 1.72$^b$ %& 8128
        \\ 
        TZ0 & \Block{4-1}{Zero} & \Block{2-1}{Thermal} & Z$_{sol}$ & 0.43 & 4.38e-4 & 0.0013 & 0.445 & 0.99$^b$ & 15.6$^b$% & 1035 
        \\
        T0.01Z0 & & & 0.01 Z$_{sol}$ & 0.16 & 4.71e-4 & 0.0015 & 0.563 & 1.25$^b$ & 19.7$^b$% & 4920 
        \\ 
        CZ0 & & \Block{2-1}{Circular} & Z$_{sol}$ & 0.35 & 1.9e-4 & 0.001 & 0.17 & 0.38$^b$ & 5.96$^b$ %& 1058 
        \\ 
        C0.01Z0 & & & 0.01 Z$_{sol}$ & 0.16 & 1.56e-4 & 0.0015 & 0.187 & 0.41$^b$ & 5.56$^a$ %& 5905
        \\ 
    \end{NiceTabular}
    \begin{flushleft}
    $^a$ Not enough wide BBHs for statistical analysis.
    $^b$ Extrapolated from the low-density, low velocity dispersion case.
    \end{flushleft}
    \caption{The properties of the wide binaries models studied, along with wide BBH fractions and BBH GW merger rates. Models differ in the natal kick (NK) model used in each simulation; initial eccentricity distribution; and metallicity. $f_{wide}$ is the fraction of wide BBHs, with SMA larger than $10^3$ AU, out of all BBHs, while $f_{BBH}$ is the fraction of wide eccentric BBHs out of the whole ensemble. The merger ratios are averaged per NK model. The rates are calculated by obtaining merger fractions from detailed flyby simulations for only single Solar-neighborhood like environment and then extrapolating the rate for the whole galaxy. We considered the higher density environment of the inner part of a spiral galaxy and the higher density and velocity dispersion of elliptical galaxies only for one model, CZB, and then extrapolated the fiducial Solar-neighborhood model of the corresponding NK case, by multiplying with the same factors as derived from the CZB model.}
    \label{tab:grid}
\end{table*}

\section{Results} \label{sec:results} 
\subsection{Formation of wide BBH and their properties}
In this section we describe the characteristics of the BBH populations as generated by COMPAS, focusing on the progenitor wide-binary ($>1000$ AU) populations, typically ignored in models for isolated evolution channels of GW sources. We focus on BBHs because they are the heaviest and most common double compact objects among already observed GW events, however similar analysis can be performed on neutron star binaries and the channels for black hole–neutron star binaries. These will be considered in future studies.

We note that in models with full natal kicks, very few (none for the Solar metallicity cases) wide binaries were formed, and the statistics are poor, providing limited characterization of the wide binary properties, and effectively producing no (or too few as to measure with the current statistics) BBH mergers from wide binaries. Given the resulting low numbers, the detailed characterization of these models is of lesser importance.

%and the properties of the merging binaries, as to be compared with current and future inferred properties of BBH mergers in VLK.

The formation of wide BBHs from shorter period binaries strongly depend on the natal kicks given to the BHs; larger kicks are generally more disruptive and give rise to smaller fractions of wide binaries. Nevertheless, since our initial population of binaries input to COMPAS consisted of {\bf no} wide binaries, the wide binary population progenitors were all produced due to stellar evolution, mass loss and NKs which evolved shorter period binaries into wide binaries. In other words, significant populations of wide BBHs can form from shorter period binaries, besides the cases of very large NKs (Full models), which give rise to very small (if any) fractions of wide binaries.

Additional wide binaries could possibly form either through direct star formation or through capture \citep[e.g.][]{2012ApJ...750...83P}
and further increase the rates considered here.

In Figures~\ref{fig:sma}-\ref{fig:q} we show the orbital separation for all BBHs resulting for the COMPAS population synthesis modeling, and the eccentricity and mass properties of all the formed {\emph wide} BBHs.

\subsubsection{Semi-major axes}
The
fractions of systems with SMA larger than $10^3$ AU (Fig. \ref{fig:sma} (see also Table \ref{tab:grid}) ranges between $0.1-0.4$ in the various models \cite[comparable to the fraction assumed in, e.g.,][]{2019MNRAS.486.4098I,2012ApJ...750...83P}. By multiplying these fractions with the fraction of MS binaries that become BBHs, we obtained the fraction of wide eccentric BBHs, $f_{BBH}$, shown in Table~\ref{tab:grid} per NK model. 

\subsubsection{Eccentricities}
In some of the models the eccentricity distributions we find for the wide BBH populations models follow  approximately a thermal distribution  (Fig.~\ref{fig:ecc}), as assumed in \cite{2019ApJ...887L..36M},  while in others, the resulting distribution is slightly superthermal. However, in models with initial zero eccentricity for the initial binaries (which are likely less realistic) the resulting wide binaries eccentricity distribution is quite different. 

\subsubsection{Masses}
Figures \ref{fig:mass}-\ref{fig:q}, show the total mass of the wide BBHs  and their mass ratios distributions. As can be seen the masses and mass-ratios are very similar for the $<1000$ AU binaries and the wide binaries populations in zero-kick models, while in fallback models and full kick models, the kicks preferentially unbind lower mass BBHs, rather than transforming them into wider binaries. The latter have lower escape velocities (and in the fallback models lower-mass BHs effectively receive higher natal kicks), explaining this preference.
This introduces a bias towards more massive BBHs, which then propagates to the masses and mas-ratios of merging BBHs discussed below.

\subsection{The rates and properties of GW-merging wide BBHs}

\subsubsection{Merger rate}
The calculation of the GW merger rate includes the fraction of wide binaries formed due to stellar evolution as obtained from the COMPAS population results, which is then multiplied by the fraction of wide binaries that merged due to flyby perturbations, as found in our perturbations simulations. The total inferred rate is then calculated by accounting for the typical total mass in galaxies of different types and the number density of galaxies of a specific type (here only divided between spirals and ellipticals). 

The averaged merger fraction for each model is shown in Table~\ref{tab:grid}, as well as averaged merger rates; we find that the numerical results are generally consistent with \cite{2016MNRAS.458.4188M} results in the comparable NK model (zero NK). 

Given the computational costs, we ran detailed flyby simulations only for a single Solar-neighborhood like environment and extrapolated the rate for the whole galaxy. Only for one model, CZB, did we consider the higher density environment of the inner part of a spiral galaxy and the higher density and velocity dispersion of elliptical galaxies. The higher density region give rise to $\sim2.2$ times higher rate, and the elliptical galaxy environment give rise to $\sim3$ times higher rate, but the number density of ellipticals is $\approx$10 times higher, and together one gets a $\sim30$ times higher rate. We expect the merger rates in other NK models to scale in a similar manner, and show the expected resulting rates for the other models in Table \ref{tab:grid}. These are marked with a $\sim$ as to highlight that these rates do not arise from a full flyby simulation calculation, but are based on extrapolating the fiducial Solar-neighborhood model of the corresponding NK case, by multiplying with the same factors as derived from the CZB model.  

Overall, we find that the wide-binary origin for BBH GW mergers is dominated by the contribution from elliptical galaxies. Considering their contribution we find potential rate in the range $1-20$ Gpc$^{-3}$ yr$^{-1}$ for models with no NKs or fallback models for NKs (see ellipticals column in Table~\ref{tab:grid}), which generally overlaps the observationally inferred rate from aLIGO/VIRGO detections. As expected the rates are strongly dependent on the natal kicks imparted to BHs at birth. 

We find slightly larger merger rates in models of binaries with smaller metallicity. Note, however, that ellipticals, dominating the rates are typically high metallicity environments. 

%\begin{table}
%    \centering
%    \begin{tabular}{|l|c|c|} 
%    \hline
%        Model & Distribution & R-square \\ \hline \hline \hspace{0.5cm}
%        T$^\dagger$, $Z_\odot$, zero & $y\propto x^{-2.34\pm 0.09}$ & 0.98 \\ \hline
%        T, 0.01$Z_\odot$, zero & $y\propto x^{-2.3\pm 0.1}$ & 0.977 \\ \hline
%        T, $Z_\odot$, fallback & $y\propto x^{-2.9\pm 0.3}$ & 0.96 \\ \hline
%        T, 0.01$Z_\odot$, fallback & $y\propto x^{-3.6\pm 0.3}$ & 0.958 \\ \hline
%        C$^\ddagger$, $Z_\odot$, zero & $y\propto x^{-2.39\pm 0.09}$ & 0.98 \\ \hline
%        C, 0.01$Z_\odot$, zero & $y\propto x^{-2.5\pm 0.1}$ & 0.972 \\ \hline
%        C, $Z_\odot$, fallback & $y\propto e^{x(-5\pm 0.4)\times 10^{-4}}$ & 0.99 \\ \hline
%        C, 0.01$Z_\odot$, fallback & $y\propto x^{-3.6\pm 0.3}$ & 0.96 \\ \hline
%    \end{tabular}
%    \caption{$^\dagger/^\ddagger$ Initial eccentricity distribution: Thermal/Circular. The metallicity listed is the initial metallicity. Kick velocity models: zero/fallback}
%    \label{tab:my_label}
%\end{table}

%In order to find the merger fractions we implemented our flyby simulation on each wide eccentric BBH outputted from COMPAS. That was repeated for $N(a,e) = \mathcal{O}(10^3-10^4)$ to obtain the merger fraction per BBH system, $f_m$. The averaged merger ratio per NK model (see Table~\ref{tab:grid}) was then obtained by averaging over $n_{wide}$ specific merger fractions.

\subsubsection{Chirp mass and mass ratios of merging BBH} 
In \cite{2019ApJ...887L..36M}, we suggested that the wide-binary flyby channel produces merger rates that increase with the total binary mass, and that equal mass components of wide BBHs are more likely to merge through the proposed mechanism. As discussed above, formation of wide-binaries from shorter period binaries can introduce a bias towards higher mass and higher mass-ratios for wide BBHs. The flyby perturbations that drive BBH mergers in this channel typically introduce slight preference towards higher mass binaries, as suggested in our previous paper.

\begin{figure*}
\centering
\includegraphics[scale=0.55,trim = 0.52cm 0.1cm 1cm 0.5cm,clip]{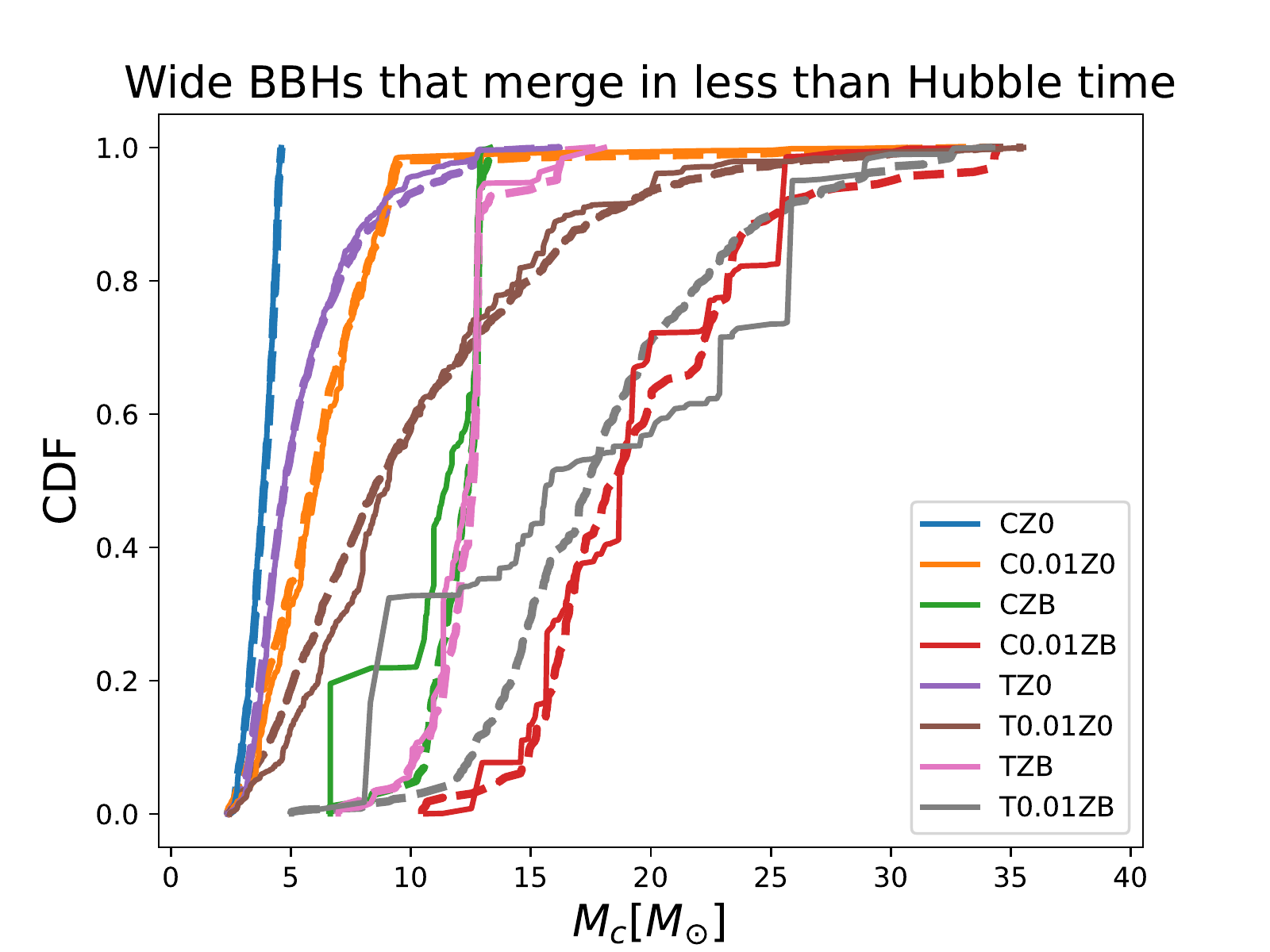}
\includegraphics[scale=0.55,trim = 0.52cm 0.1cm 1cm 0.5cm,clip]{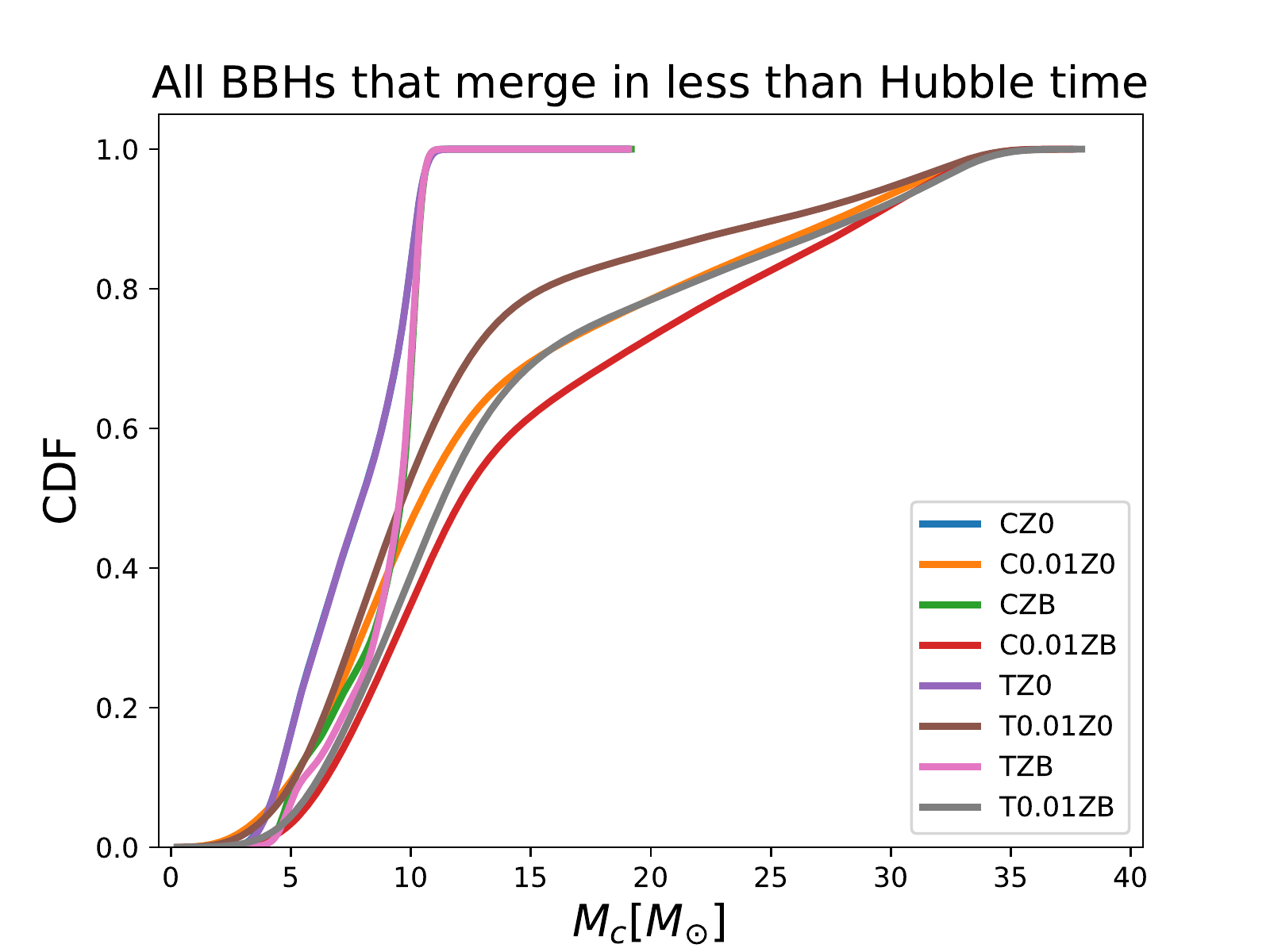}
\caption{Left: The chirp mass distributions for the wide eccentric BBH populations that merged in less than Hubble time due to flybys in the field, considering different natal-kick models. Solid lines correspond to distributions that are weighted by each system merger fraction $f_m$, while dashed lines correspond to the mass distributions assuming all $f_m=1.0$. Models of low NK velocities show preference towards GW sources from more massive binaries, while for zero kick models we obtain lighter chirp masses. Right: for comparison, we show the chirp mass distributions of the BBH merger populations generated by COMPAS from the isolated binary evolution channel (i.e. not originating from wide binaries) that merged in less than Hubble time. %\color{red} The flyby mechanism differentiates between a universe in which BHs receive no kicks at birth and a universe in which BHs are given low kick velocities at birth. \color{black}
} \label{fig:mchirp}
\end{figure*}

\begin{figure*}
\includegraphics[scale=0.45,trim = 0.6cm 0.1cm 0.8cm 0.2cm,clip]{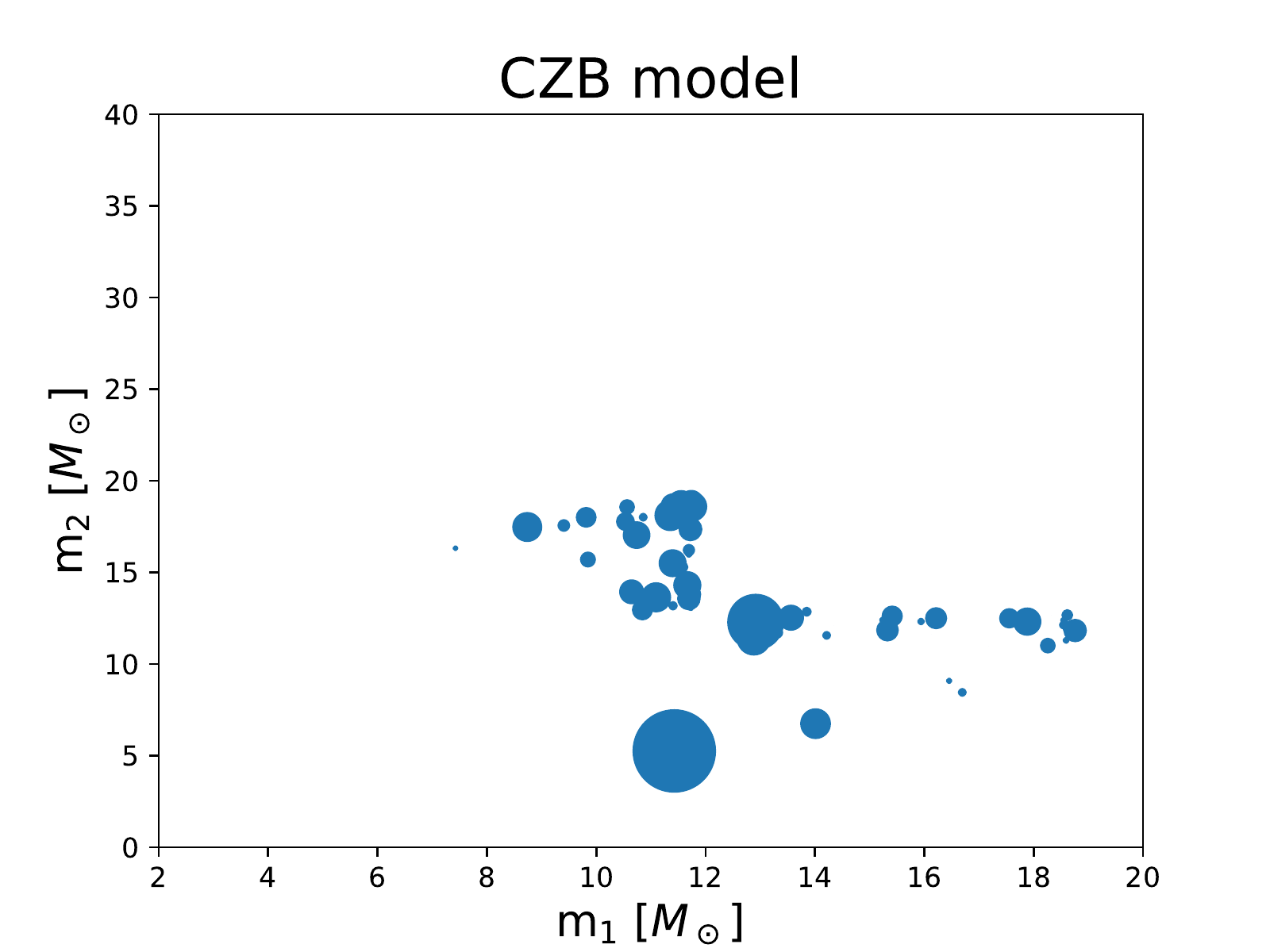}  \includegraphics[scale=0.45,trim = 0.6cm 0.1cm 0.8cm 0.2cm,clip]{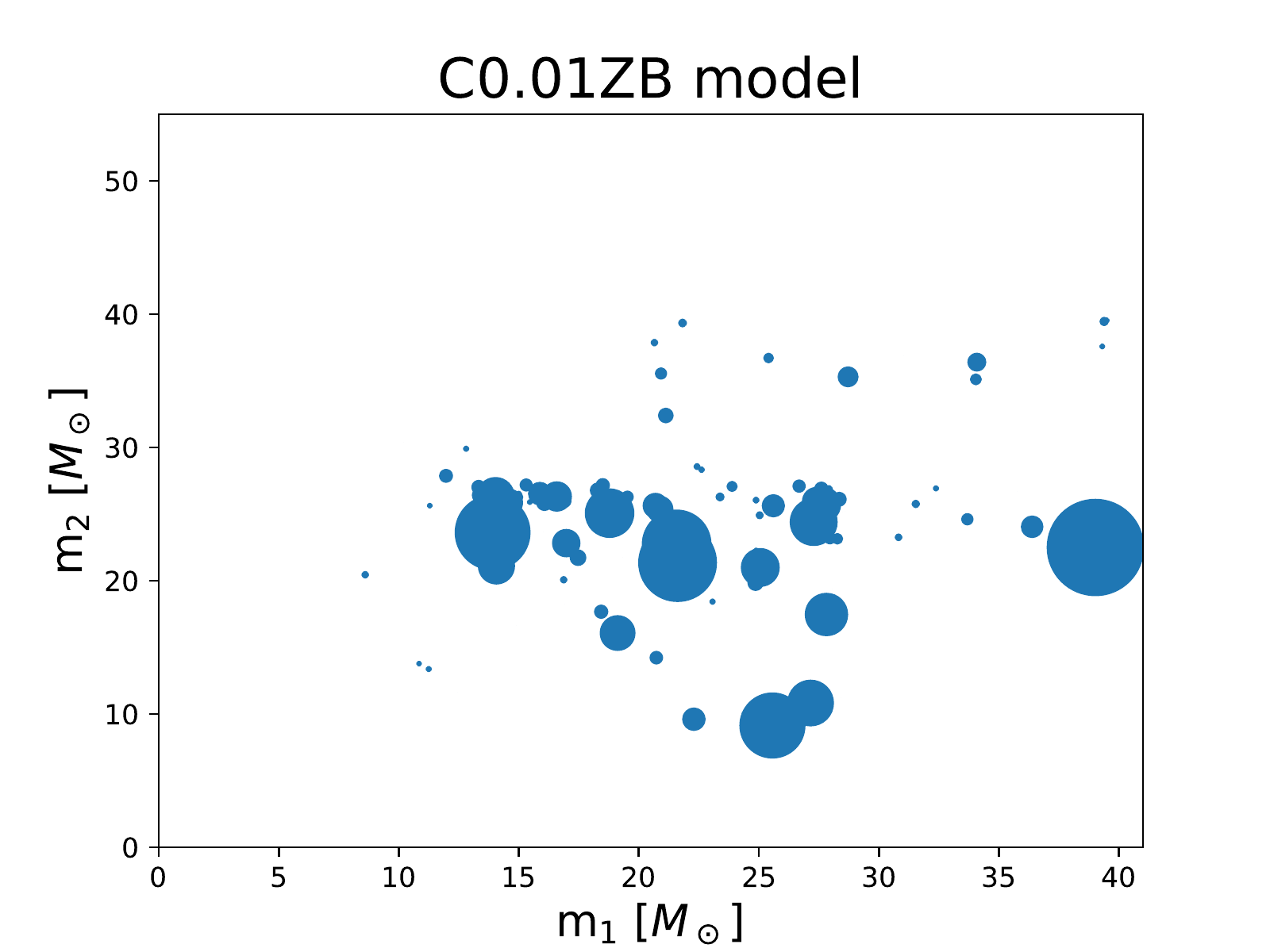}
\includegraphics[scale=0.45,trim = 0.6cm 0.1cm 0.8cm 0.2cm,clip]{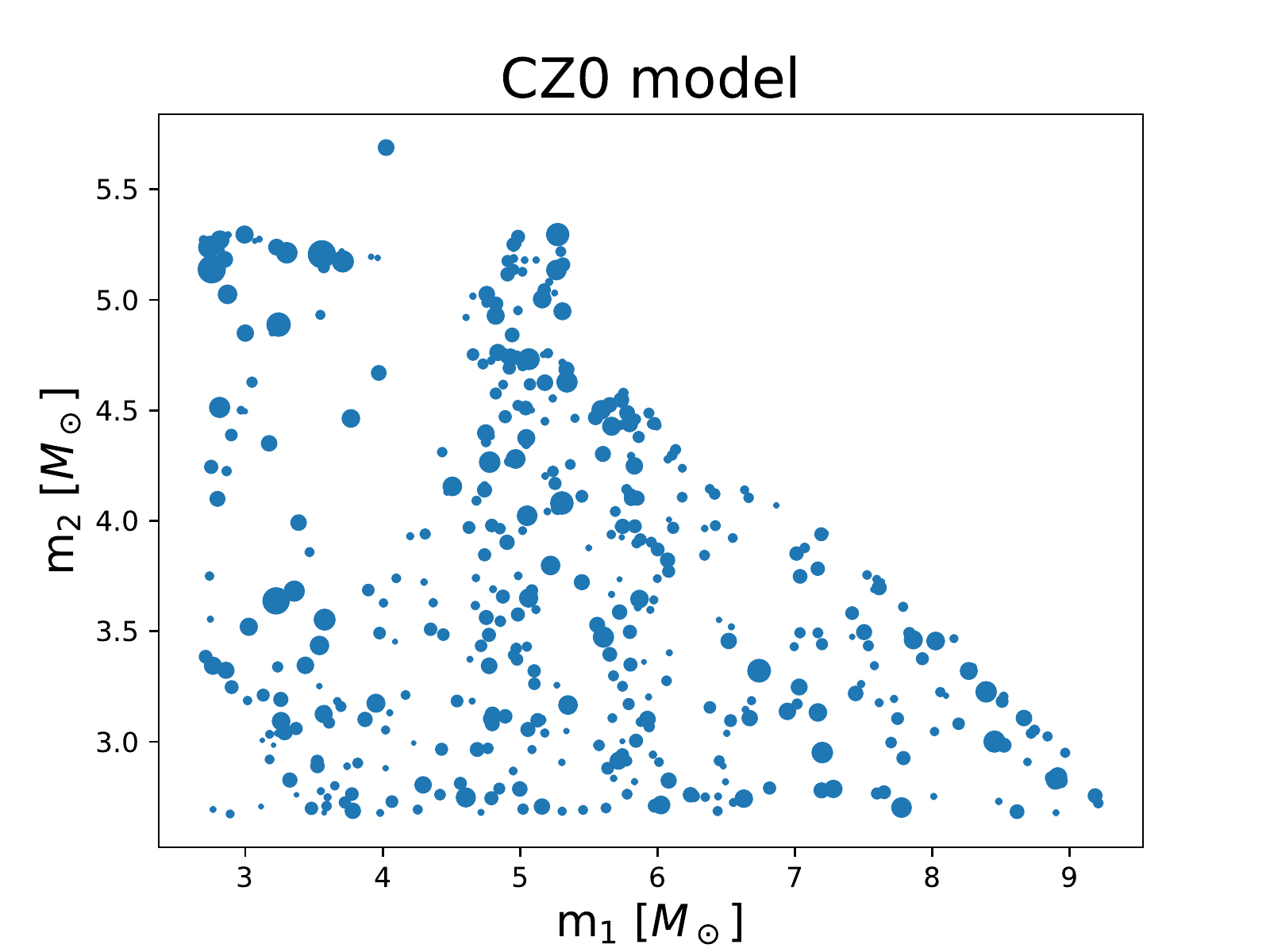}
\includegraphics[scale=0.45,trim = 0.6cm 0.1cm 0.8cm 0.2cm,clip]{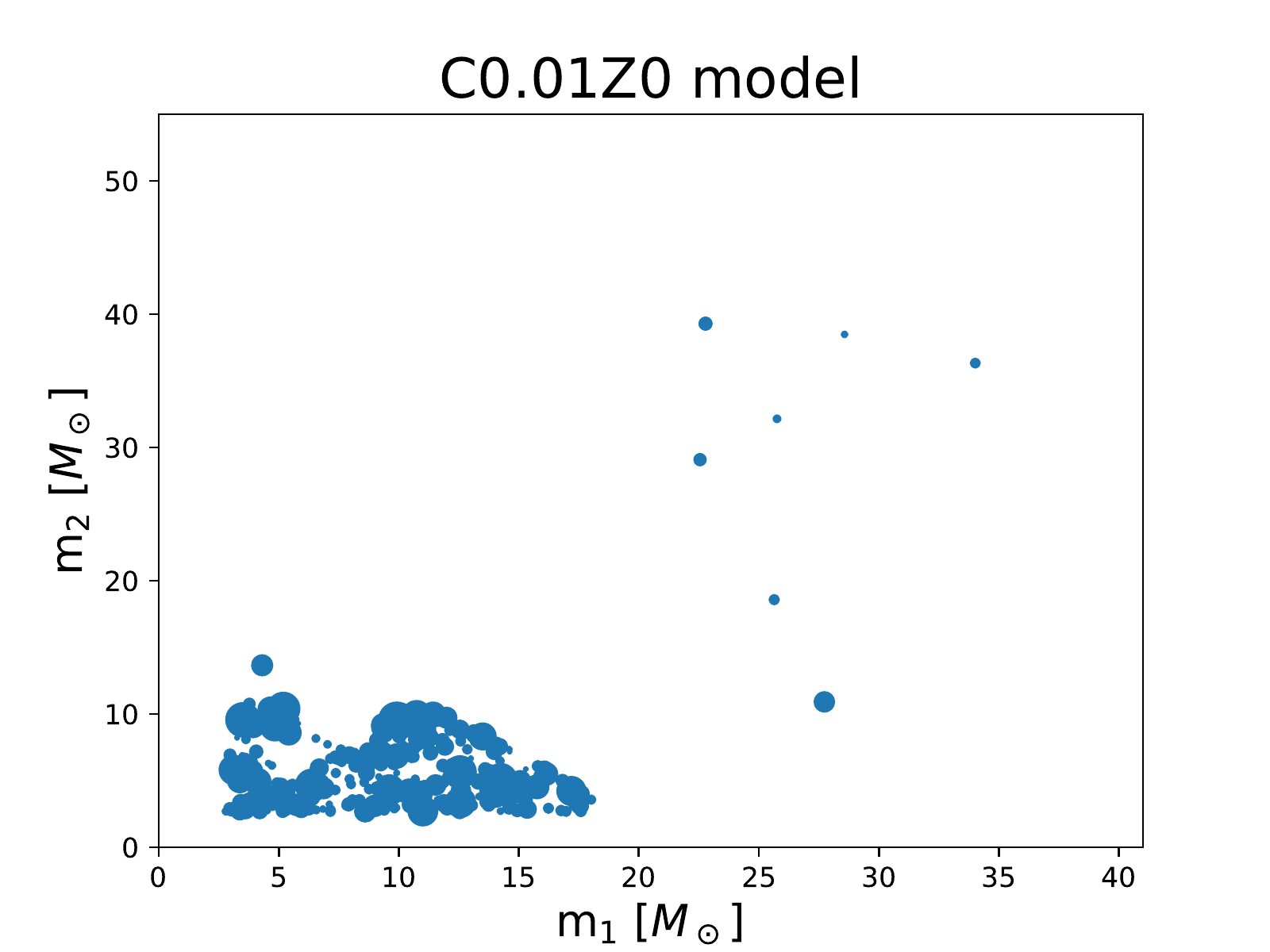} \\ \includegraphics[scale=0.45,trim = 0.6cm 0.1cm 0.8cm 0.2cm,clip]{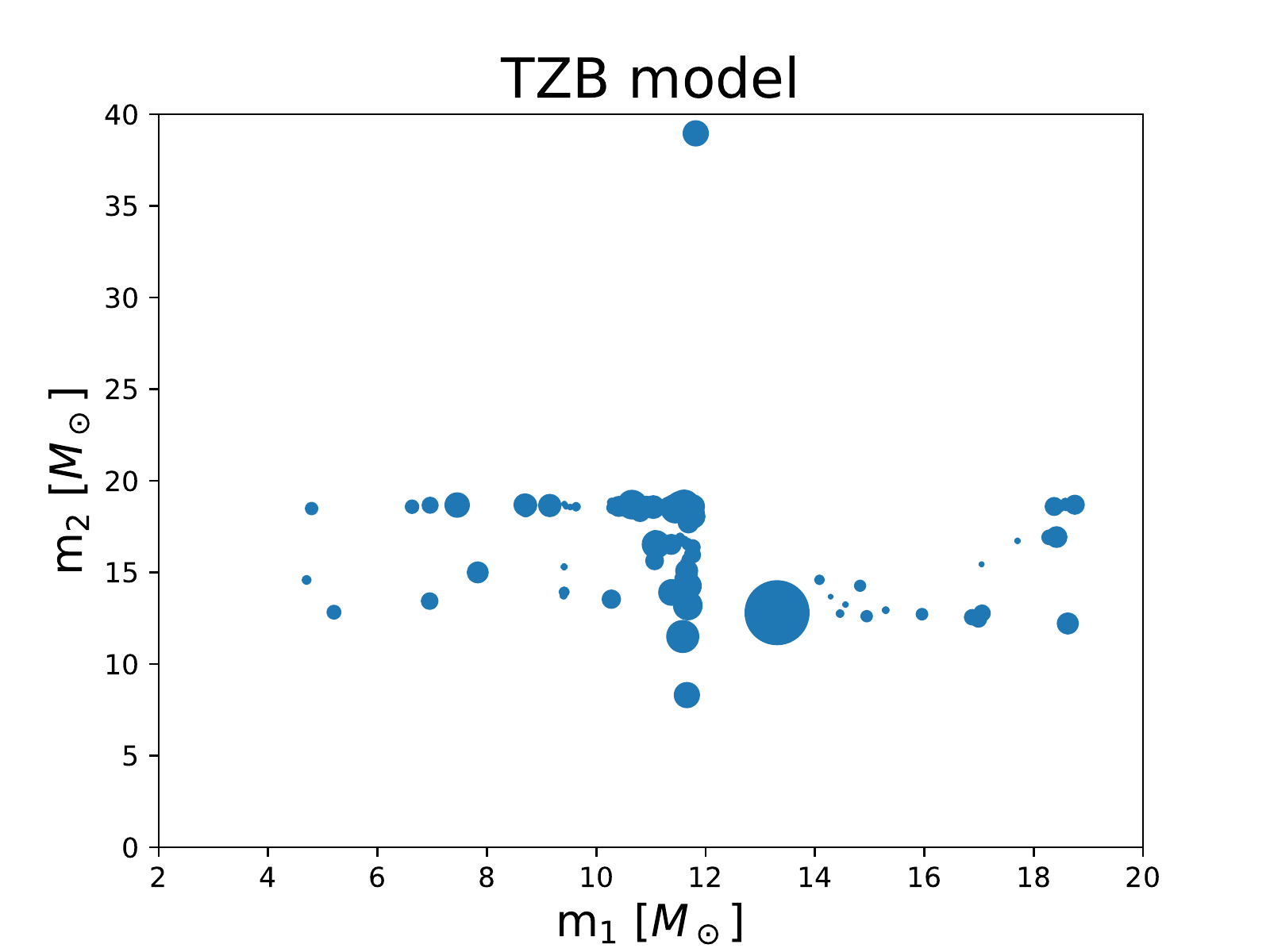}  \includegraphics[scale=0.45,trim = 0.6cm 0.1cm 0.8cm 0.2cm,clip]{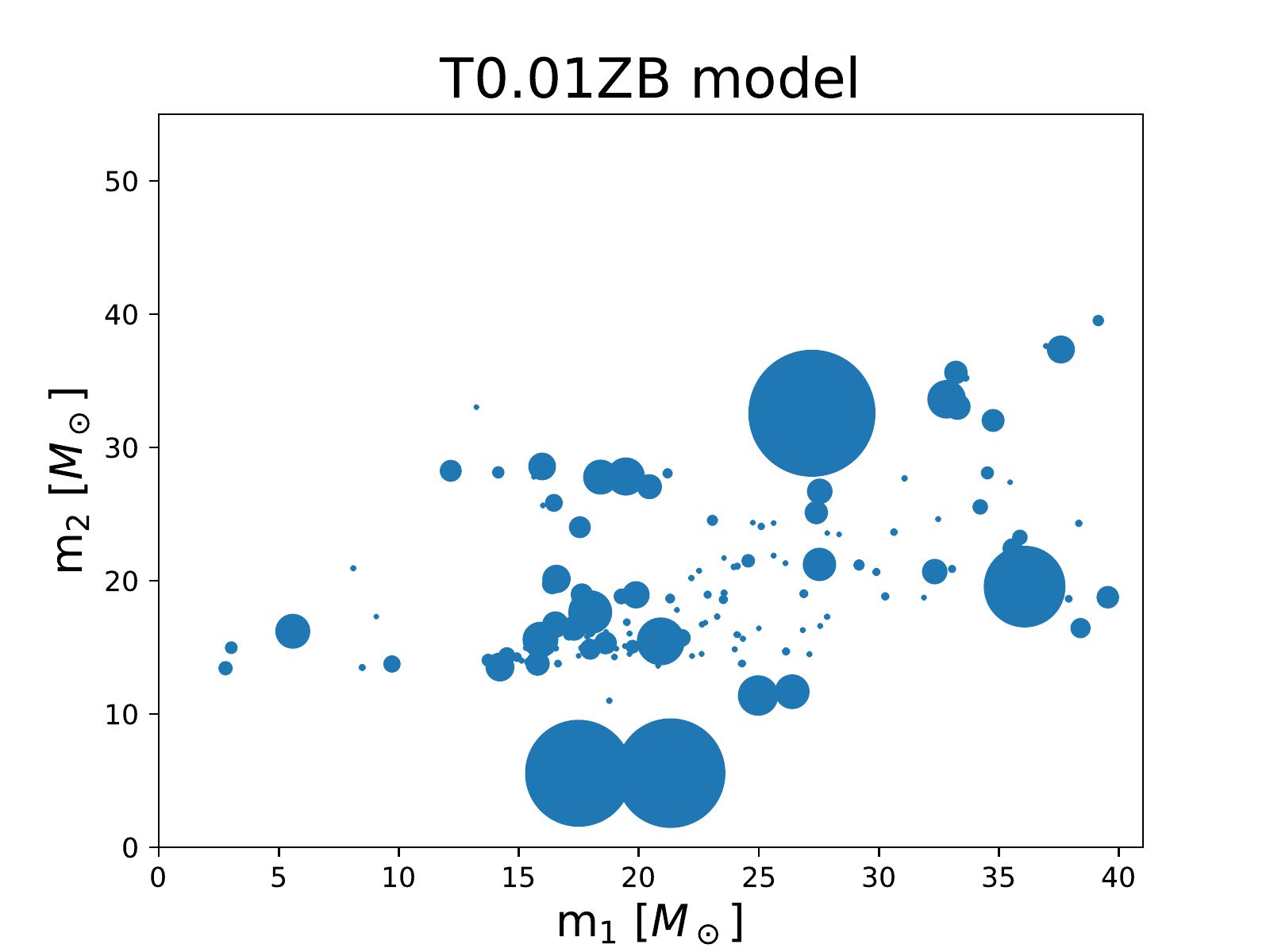}
\includegraphics[scale=0.45,trim = 0.6cm 0.1cm 0.8cm 0.2cm,clip]{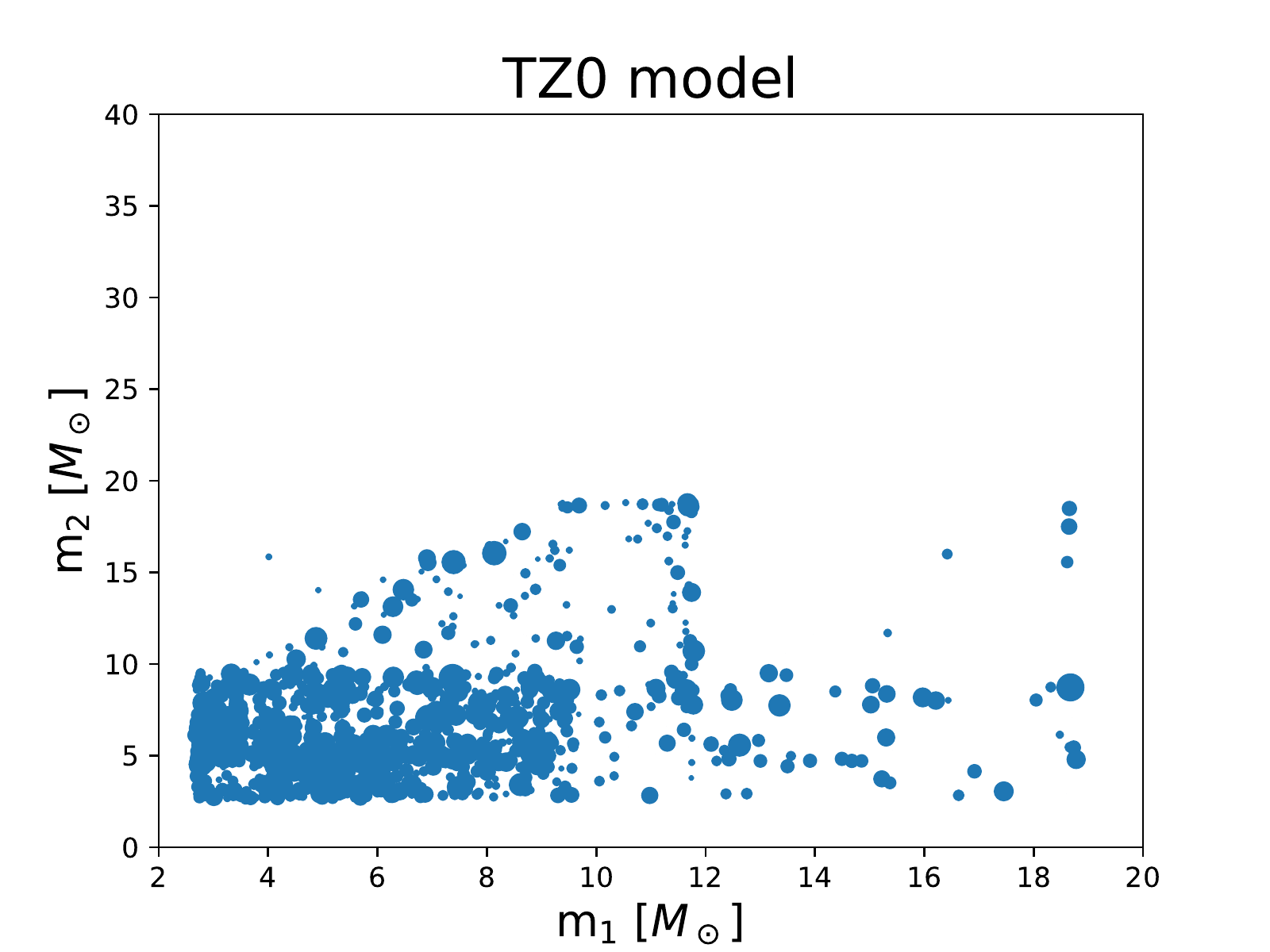}
\includegraphics[scale=0.45,trim = 0.6cm 0.1cm 0.8cm 0.2cm,clip]{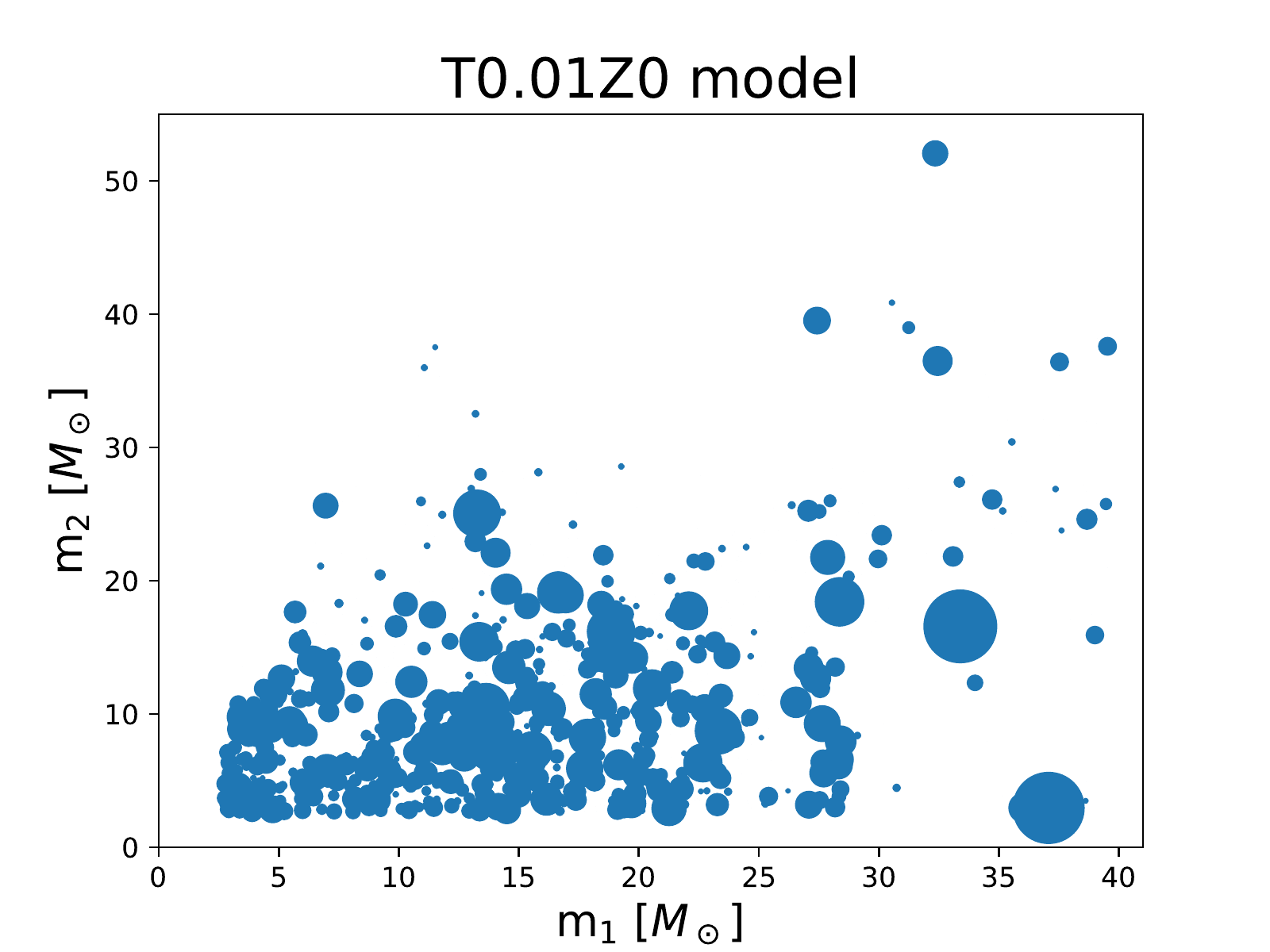}
\caption{Masses of the wide eccentric BBHs' components, once the binaries are formed in COMPAS, considering different natal-kick models. The size of the points encapsulates the binaries' merger fraction $f_m$ due to flybys in the field. No distinct correlation is found between the merger fraction of perturbed wide binaries in the field and the binaries' mass or mass ratio. There are no massive binaries in the relatively high metallicity models due to high mass loss rate through line-driven winds during the binaries formation. 
} \label{fig:m1m2}
\end{figure*}

Figure \ref{fig:mchirp}-\ref{fig:m1m2} appear to be consistent with these expectations. The left panel in Fig. \ref{fig:mchirp} shows the cumulative distributions of chirp masses of merged wide BBHs (solid lines), is preferentially biased towards more massive binaries, compared with all the wide BBHs modeled through flyby perturbations. The right panel shows for comparison the resulting chirp mass distribution of merging BBHs in the isolated evolution channel from the same initial populations modeled in COMPAS. As can be seen the wide-binary channel produce systematically higher chirp masses. The difference, however,  mostly arises from the initial formation of the wide binaries through kicks, with far lesser effect of the flybys in driving the mass preference.

%Figure~\ref{fig:mchirp} shows the cumulative DF of chirp masses per NK model; solid lines are weighted by the merger ratio $f_m$ of each system in a model. Models of low NK velocities show preference towards GW sources from more massive binaries, while for zero kick models we obtain lighter chirp masses. 

%\subsubsection{Initial (MS binary) SMAs vs wide BBH SMAs}
%We expect that if wide BBHs originate from initial binaries with small separations, they would have correlated spins.
%\subsubsection{The properties of wide binaries}
In Figure~\ref{fig:m1m2} we show the masses of the components of wide eccentric BBHs which eventually merged, weighted by the merger fraction $f_m$ (shown as larger circles), for each of the NK models. No distinct correlation is found between the merger fraction of perturbed wide binaries in the field and the binaries' mass or mass ratio. We regard the absence of massive binaries with solar metallicity as stemming from large mass loss rate, as it is known that metallicity significantly impacts the rate of mass loss through line-driven winds during the binaries formation \citep{2010ApJ...714.1217B,2018MNRAS.480.2011G}.

\subsubsection{Delay time distribution}
\begin{figure*}
\centering
\includegraphics[scale=0.55,trim = 0.1cm 0cm 0.5cm 0.5cm,clip]{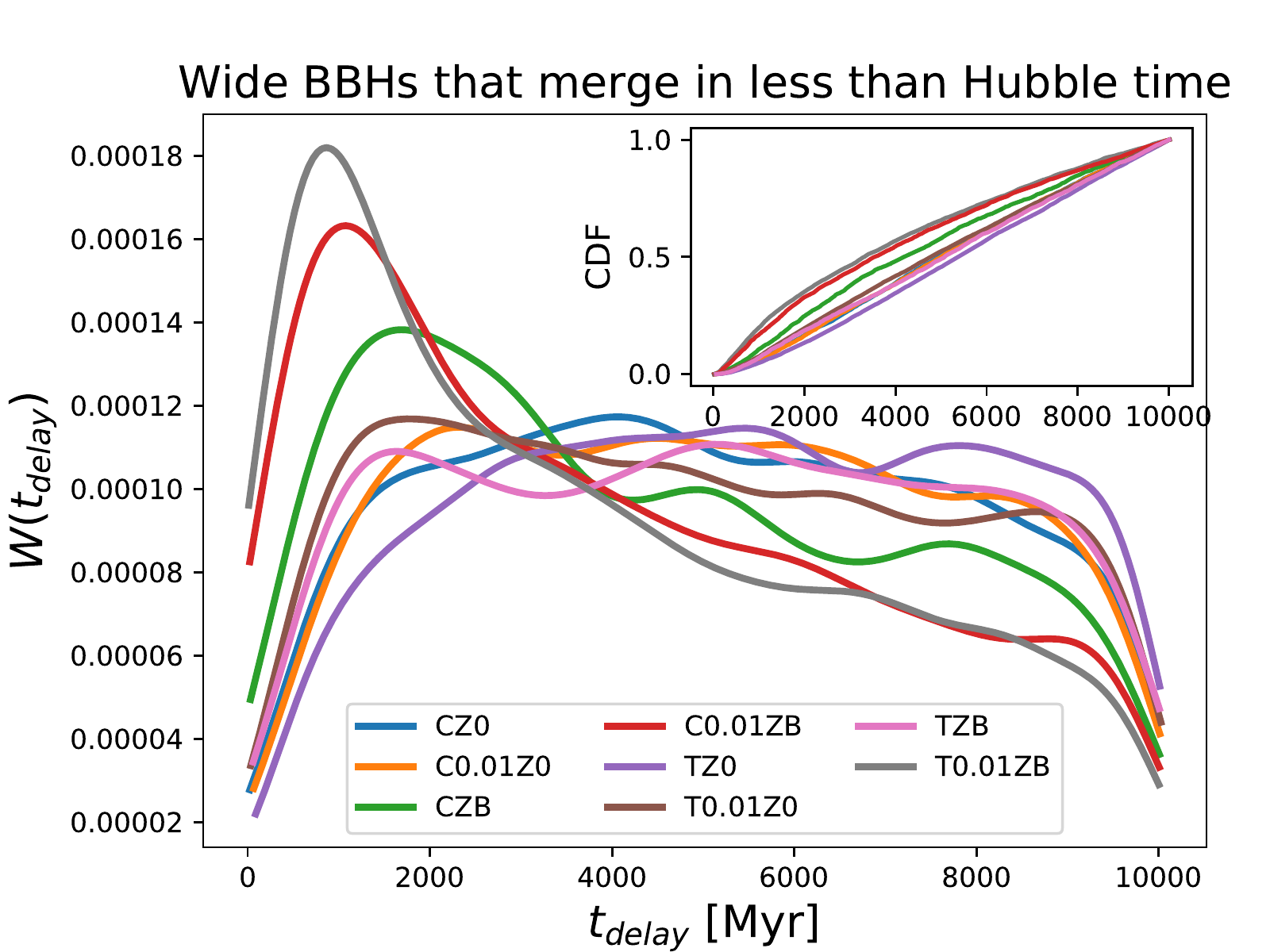}
\includegraphics[scale=0.55,trim = 0.2cm 0cm 0.7cm 0.5cm,clip]{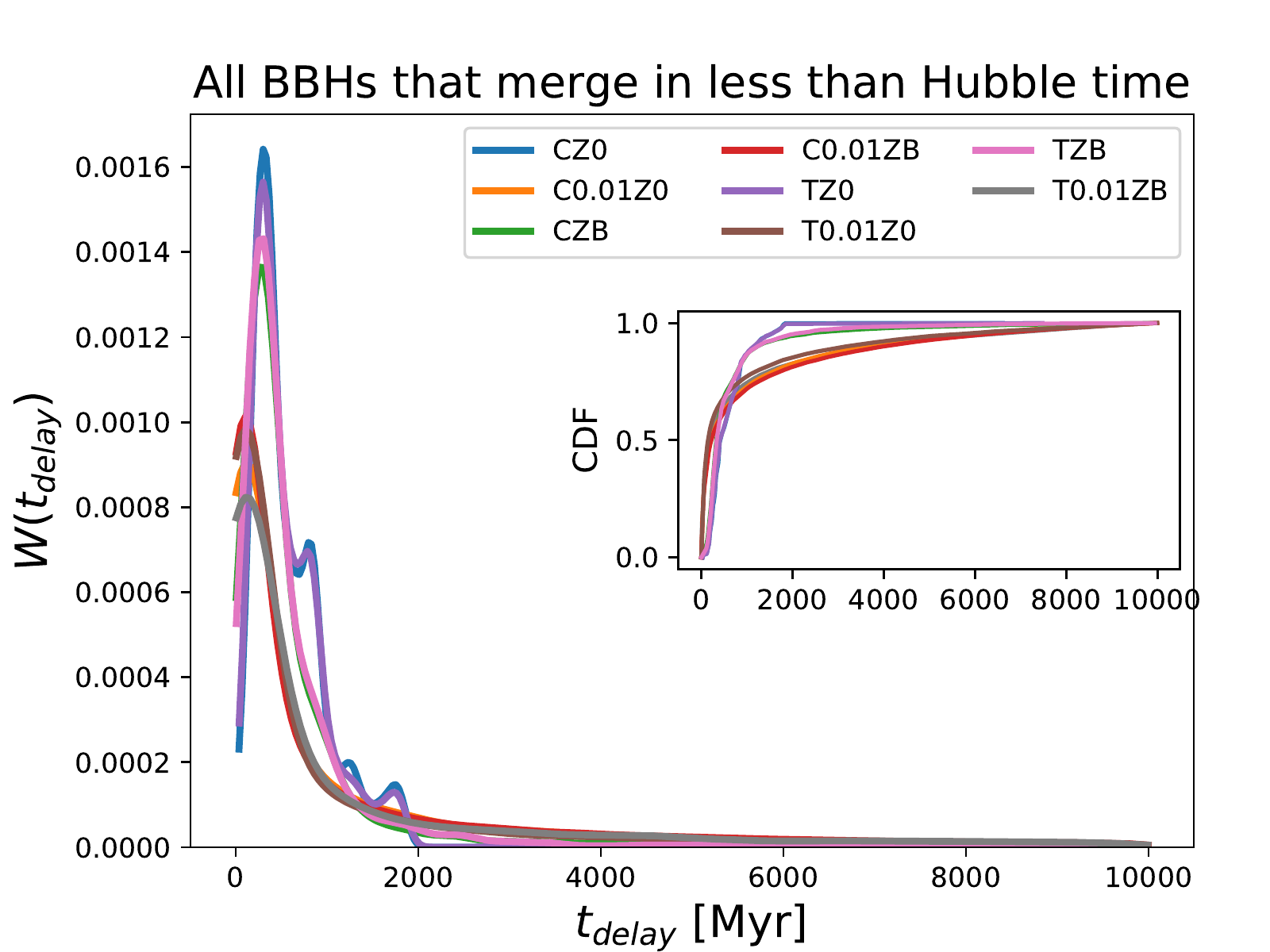}
\caption{Delay time distributions, i.e. merger time since star formation. Left: wide eccentric BBH populations that merged in less than Hubble time due to flybys in the field, considering different natal-kick models. The distributions are generally uniform, with some preference for early mergers in the first 2 Gyrs for some of the models. Right: for comparison, the DTD for mergers of non-wide BBHs from the isolated binary evolution channel generated by COMPAS. These distributions generally follow $t_{delay}(t)\propto t^{-1}$, where most of the currently observable GW sources arise from relatively recent star-formation (last 2 Gyrs).
}\label{fig:dtd}\end{figure*}

The delay time distributions (DTDs) of all simulated merging BBH populations 
%deviate from a uniform distribution by factors smaller than $\sim 2$ 
are shown in Fig.~\ref{fig:dtd}.

The distributions are generally uniform, with some preference for early mergers in the first 2 Gyrs for some of the models. In comparison, the DTD for mergers of non-wide BBHs from the isolated binary evolution channel generated by COMPAS is strongly preferential for short DTD of approximately 0.5-2 Gyrs \citep[e.g.]{Britt2021}. The latter generally follows $t_{delay}(t)\propto t^{-1}$, where most of the currently observable GW sources would therefore arise from relatively recent star-formation (last 2 Gyrs), and be dominated by spiral host galaxies. In contrast, the wide BBH channel predicts a significant contribution from all delays, with a significant contribution from elliptical galaxies. 

\subsubsection{BBH merger eccentricities}
In order to calculate eccentricities at LIGO's frequency band we used the following scheme: %To determine when a binary crosses into the LIGO/Virgo band, we must know the eccentricity and semi-major axis when that binary’s GW frequency passes 10 Hz. 
the dominant frequency at which an eccentric binary emits GWs can be approximated as %We calculate the approximate gravitational peak frequency following
\citep{2003ApJ...598..419W}
\be f_{peak}(a,e)=\dfrac{1}{\pi}\sqrt{\frac{G(m_1+m_2)}{a^3}}\frac{(1+e)^{1.1954}}{(1-e^2)^{1.5}}, \ee
where $a$ and $e$ are the semi-major axis and eccentricity, respectively. To determine when a binary enters the LIGO/Virgo band, we integrate the orbit averaged equations for the evolution of $a$ and $e$ from \cite{PhysRev.136.B1224} until $f_{peak} = 10$ Hz \cite[e.g.,][]{2018PhRvD..98l3005R}. %In LISA we follow the orbit averaged equations until $f_{peak} = 0.5$ Hz \citep{2021arXiv210316030R}. 

\begin{figure}
\centering
\includegraphics[scale=0.55,trim = 0.5cm 0.1cm 1cm 0.5cm,clip]{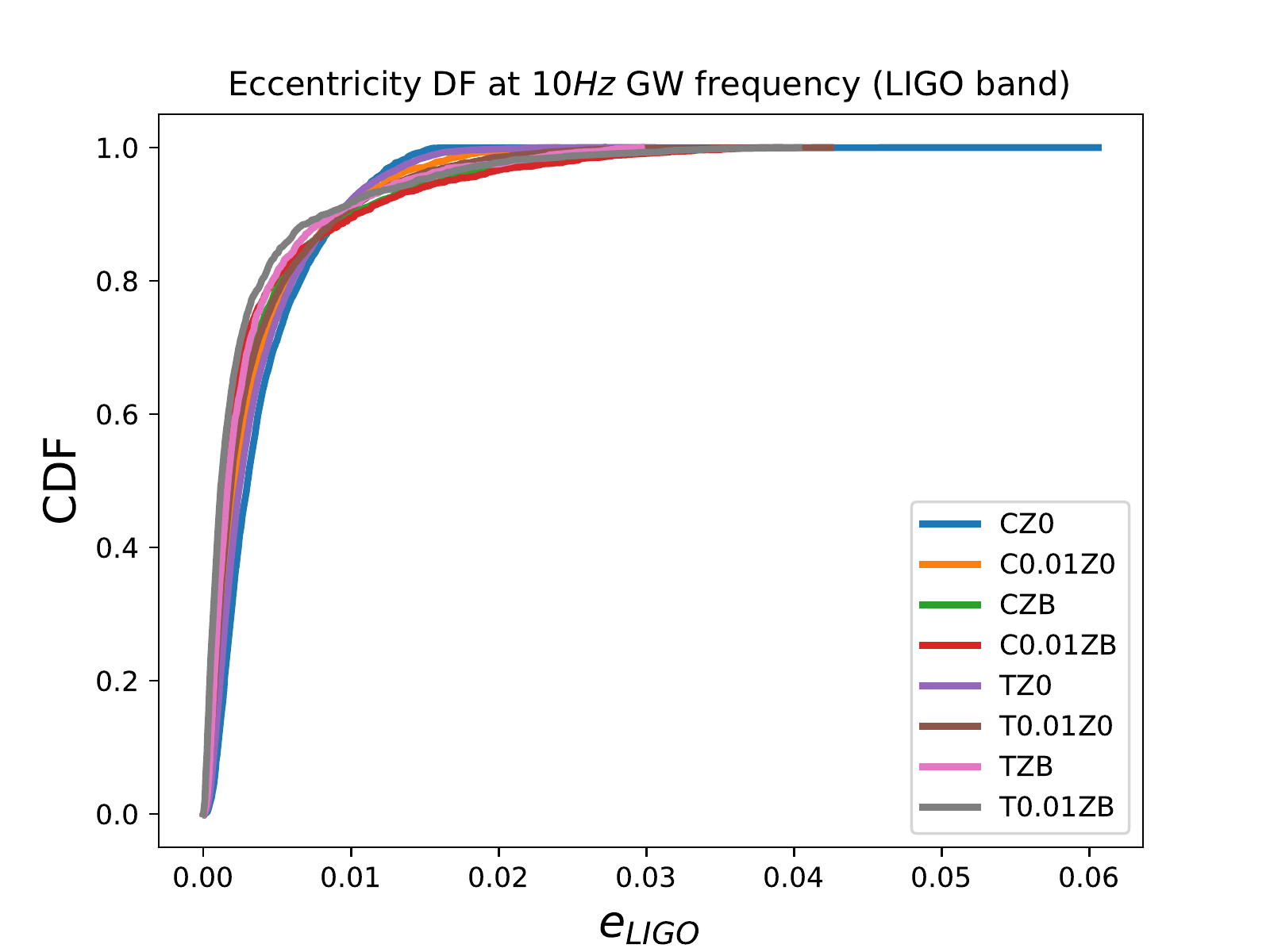}
\caption{The eccentricity distribution at 10 Hz GW frequency of the initially wide and eccentric BBH populations, considering different natal-kick models. As expected, the orbits of the BBHs circularize by the time they reach the aLIGO band and to not produce any eccentric binaries at these frequencies, in contrast with some of the dynamical channels. Eccentric mergers could, however, arise from wide-triples as we discussed elsewhere.
} \label{fig:eligo}
\end{figure}

In Fig.~\ref{fig:eligo} %and Fig.~\ref{fig:elisa} 
we present the normalized eccentricity distribution at $10$ Hz %and $0.2$ Hz
, i.e. in the LIGO/Virgo band, for each of the different ensembles described in Table \ref{tab:grid}.% weighted by the inner SMA distribution.
We report that none of the mergers are eccentric in the aLIGO band. We should note that some fractions of eccentric mergers would be produced through flyby perturbations of wide triples \cite{mic+20} not explored here.

\section{Discussion} \label{sec:sum} 
%Our analysis, using population synthesis models shows that wide $>1000$ AU BBHs can robustly form from shorter period binaries, and that even moderate natal kicks, e.g. as expected from fallback models give rise to large population of BBHs. We then studied the flyby perturbations evolution of such wide BBHs and the production and WG mergers from highly eccentric wide BBHs. We characterized the rates and properties of such mergers.
In our study we find that the merger rates from the wide binary flyby evolutionary channel proposed in \cite{2019ApJ...887L..36M} depends on the natal kicks given to BHs, but it is not more sensitive to this than the isolated evolution of dynamical channels studied by others.  We find that wide binaries progenitors are robustly produced from shorter period binaries due to stellar evolution - mass-loss and natal kicks, even if such wide binaries did not initially form.
Although the fractions of wide BBHs in populations that experienced low natal kicks are smaller than wide BBH fractions in populations that experienced zero natal kick, they are still significant, and therefore the wide-binary channel is not fine-tuned for only zero NKs.

We find that full NK models fail to produce the GW rates (far too low) in the wide binaries channel studied here. However, we should note that full NK models for BHs are effectively excluded in {\bf all} of the currently suggested formation channels, including all the isolated and dynamical channels studied in the literature to data (to the best of our knowledge), i.e. this sensitivity is not unique to the wide binary channel and no current GW-merger channel can reproduce the inferred BBH merger rates assuming full NK for BHs.

%Moreover, according to our results (see Table~\ref{tab:grid}), wide BBHs simply don't form if BHs are given full momentum kick at birth. We conclude that adapting no-kick assumptions for BHs, as done by other potentially successful scenarios \cite[e.g.,][]{}, suggests that the wide-binary channel can give rise to non-negligible production rate of GW sources from BBH mergers of perturbed wide binaries.

The wide binary channel gives rise to speciﬁc characteristics of BBH mergers, which together can provide a distinct signature for this channel. The most significant signature is a far stronger preference for long delay time distribution compared with isolated binary evolution models, which would therefore give rise to a significant contribution of GW sources from elliptical host galaxies. 

Since wide BBHs formed from shorter period binaries are preferentially more massive, the expected chirp masses from this channel are somewhat higher than the the isolated evolution case. However, high metallicity environments give rise to less massive BHs than low-metallicity environments, which would affect the mass distribution of currently observable BBH mergers. 
The flyby model significant contribution from elliptical galaxies, which have typically higher metallicities than spiral galaxies, would give rise to low mass mergers in VLK, compared with the isolated binary evolution where VLK-observable mergers are dominated by BBHs from low-metallicity, spiral galaxies. 
The exact chirp mass-function is still not well characterized by VLK, but current data suggest a multi-modal distribution with both low-mass and high chirp-mass BBHs. 

Comparing the wide binary channel to the isolated evolution channel in terms of rates, we generally find that the isolated binary evolution models with zero natal kicks or fallback kicks produce overall merger rates which are tens up to hundred times higher than the wide-binary channel. However, most of the mergers in the isolated evolution occur on short delay times, and therefore the observable GW-merger rate in the local universe is significantly quenched (by 1-2 orders of magnitudes) compared with the total rate throughout the history of the universe (consistent with other studies of the isolated evolution channel), while the long DTDs in the wide binary channel suggest more constant rates, such that the rate in the local universe is of the same order as the rate in the past. 
We note that although we use the same population synthesis model for the formation of BBHs in the isolated evolution channel and the wide binary channel, the models have very different parameters sensitivities. In particular, the BBHs merging in the isolated evolution channel are non-negligibly sensitive to various parameterizations of the common envelope evolution and mass transfer rate, which do not affect wide binaries. For an extreme example - if common envelope is extremely inefficient, BBH merger might be significantly quenched, while the wide-binary channel would not be affected.  

Whether the spins of the merging BBH components are correlated or not is unknown, but given that the binaries likely formed from the same collapsing cloud in the model studied here, it is likely that the spins are both correlated with the orbit, and are generally prograde. The flybys are likely to somewhat smear any such spin-orbit and spin-spin correlation, and allow for more isotropic distributions, but likely still showing some correlation. Such correlations are likely smaller than expected in isolated binary evolution models, but higher than in dynamical channels, where the component BBHs are many times randomly paired.

 We found, as expected, that BBHs circularize by the time they reach the aLIGO band and to not produce any eccentric binaries at these frequencies, in contrast with some of the other dynamical channels. However, as mentioned above, flybys of wide triples \citep[][]{mic+20}, not studied here, could give rise to eccentric mergers.
 
 %Finally, in our scenario the distribution of merger time since star formation, the delay-time distribution, differs from the isolated binary channel. The latter predicts a $\propto t^{-1}$ dependence \citep{2015ApJ...806..263D} while for the most dominant NK models that contribute to the overall rate from the wide binary channel, we obtained uniform delay-time distributions.

Detecting this mechanism's distinct signal might also give insight into averaged velocity dispersion of galaxies, since the rates depend on the BBHs environment, namely the stellar density $n_*$ and the encounter velocity $v_{enc}$. We obtained an increased merger probability, by a factor of $\sim 35$, in our test case of elliptical galaxies. We therefore expect a preference for host galaxies with higher velocity dispersion. We also obtained increased merger probabilities, by a factor of $\sim 2.2$, in our test case of systems in denser areas of the galaxy. We therefore expect a preference for denser host galaxies.

\subsection{Caveats and recommendation for future research}
Overall, our results give confidence in the significance and distinguishability of the GW signal from catalyzed mergers of binaries due to flyby interactions in the field, compared with both dynamical and isolated evolution channels, albeit with some caveats.

Firstly, it is known that the star formation rate (SFR) is non-constant over the lifetime of the Universe, i.e. it changes significantly as a function of redshift \citep{2014ARA&A..52..415M}. Since SFR determines the amount of stellar binaries formed, it introduces an uncertainty in the rate of BBH formation. Furthermore, the star formation is heavily dependent on the metallicity of the star forming gas, which also changes over the lifetime of the Universe. Considering we estimated the rates by crudely integrating over number densities of either spiral or elliptical galaxies in the local Universe, and since we did our simulation on very narrow metallicity grid, our results should be taken as scale estimates only which bracket the overall contributions. 
%As theoretical metallicity-specific star formation rate models are at present still too uncertain to obtain precise calculation of the rates, 
we leave further analysis in that regard to future studies.

Another worth mentioning caveat is our assumption that merger probabilities due to flybys in the field merely weakly depend on the stellar number density; although it seems reasonable and validated in one test case we explored, more elaborate analysis needs to be done in order to precisely characterize that dependence and adjust the calculated rates accordingly. \cite{2019ApJ...887L..36M} analytically characterized that dependence for the special case of similar mass binaries with thermal eccentricities, and found that it's, at most, proportional to the square root of the stellar number density $\sqrt{n_*}$. To characterize it computationally, one should run the flybys simulation on a full grid of different stellar densities, which is highly computationally expensive. Here we only used extrapolations based on one model.

\section{Summary} \label{sec:con}
In this study we explored the formation of wide $>1000$ AU BBHs from shorter period binaries through stellar evolution and in particular mass-loss and natal kicks. We followed the evolution of such wide BBHs through flyby perturbations by stars in the field, and showed that these can give rise to BBH GW mergers of BBHs excited to very high eccentricities. 

We previously studied a simple model of single-mass wide BBH populations, with significant non-trivial assumptions regarding the frequency of such BBHs and their properties. Here we relaxed these assumptions by self-consistently exploring the formation of such BBHs from shorter period binaries through detailed population synthesis models with the COMPAS code. This also allowed us to make direct comparisons with the isolated binary evolution models making use of the same assumptions and binary modeling. 

We find that the wide binaries flyby channel for GW sources can introduce significant contribution to the GW rates, and that its sensitivity to natal kicks is comparable to the sensitivity of the isolated evolution channel. In particular, both channels (and in principle the dynamical channel too), could potentially produce a significant fraction or even the majority of GW of observed GW sources if BHs are given zero natal kicks or momentum kicks (normalized by neutron star natal kicks), while full natal kicks (i.e. the same as neutron stars) effectively quench these channels. 

We find that the wide-binary channel give rise to qualitatively different delay time distribution compared with the isolated binary evolution channel, with a relatively uniform in time delay time distribution. 
The wide-binary channel would suggest most GW sources arise from elliptical galaxies, in contrast with the preference to spiral galaxies in the isolated evolution channel.

The chirp-mass distribution, while generally  preferentially biased towards more massive binaries than the isolated binary evolution case, would eventually contribute more to low chirp-masses mergers in the local-universe. This occurs because of the far larger contribution of GW-sources from higher metallicity environments (ellipticals), in this channel. 

Regarding other properties, the mergers eccentricities are expected to be low. While the spins are not directly studied here, we expect  spin-spin and spin-orbit correlation to exist (in particular most mergers are likely to be prograde), but be less pronounced than the isolated evolution channels, and more pronounced than dynamical channels. 

Finally, our results for the robustness of the wide binary channel suggest it could produce significant contributions to other types of wide binary mergers leading to e.g. supernovae from white-dwarf mergers, and other transients. 

%\section*{Acknowledgements}
\section*{Data availability}
The data that support the findings of this study are available from the corresponding author upon reasonable request.

\bibliographystyle{mnras}
\bibliography{wideSMA}

\begin{thebibliography}{}
\makeatletter
\relax
\def\mn@urlcharsother{\let\do\@makeother \do\$\do\&\do\#\do\^\do\_\do\%\do\~}
\def\mn@doi{\begingroup\mn@urlcharsother \@ifnextchar [ {\mn@doi@}
  {\mn@doi@[]}}
\def\mn@doi@[#1]#2{\def\@tempa{#1}\ifx\@tempa\@empty \href
  {http://dx.doi.org/#2} {doi:#2}\else \href {http://dx.doi.org/#2} {#1}\fi
  \endgroup}
\def\mn@eprint#1#2{\mn@eprint@#1:#2::\@nil}
\def\mn@eprint@arXiv#1{\href {http://arxiv.org/abs/#1} {{\tt arXiv:#1}}}
\def\mn@eprint@dblp#1{\href {http://dblp.uni-trier.de/rec/bibtex/#1.xml}
  {dblp:#1}}
\def\mn@eprint@#1:#2:#3:#4\@nil{\def\@tempa {#1}\def\@tempb {#2}\def\@tempc
  {#3}\ifx \@tempc \@empty \let \@tempc \@tempb \let \@tempb \@tempa \fi \ifx
  \@tempb \@empty \def\@tempb {arXiv}\fi \@ifundefined
  {mn@eprint@\@tempb}{\@tempb:\@tempc}{\expandafter \expandafter \csname
  mn@eprint@\@tempb\endcsname \expandafter{\@tempc}}}

\bibitem[\protect\citeauthoryear{{Abt}}{{Abt}}{1983}]{1983ARA&A..21..343A}
{Abt} H.~A.,  1983, \mn@doi [\araa] {10.1146/annurev.aa.21.090183.002015},
  \href {https://ui.adsabs.harvard.edu/abs/1983ARA&A..21..343A} {21, 343}

\bibitem[\protect\citeauthoryear{{Antonini} \& {Perets}}{{Antonini} \&
  {Perets}}{2012}]{2012ApJ...757...27A}
{Antonini} F.,  {Perets} H.~B.,  2012, \mn@doi [\apj]
  {10.1088/0004-637X/757/1/27}, \href
  {https://ui.adsabs.harvard.edu/abs/2012ApJ...757...27A} {757, 27}

\bibitem[\protect\citeauthoryear{{Antonini}, {Toonen}  \& {Hamers}}{{Antonini}
  et~al.}{2017}]{2017ApJ...841...77A}
{Antonini} F.,  {Toonen} S.,   {Hamers} A.~S.,  2017, \mn@doi [\apj]
  {10.3847/1538-4357/aa6f5e}, \href
  {https://ui.adsabs.harvard.edu/abs/2017ApJ...841...77A} {841, 77}

\bibitem[\protect\citeauthoryear{{Asplund}, {Grevesse}, {Sauval}  \&
  {Scott}}{{Asplund} et~al.}{2009}]{2009ARA&A..47..481A}
{Asplund} M.,  {Grevesse} N.,  {Sauval} A.~J.,   {Scott} P.,  2009, \mn@doi
  [\araa] {10.1146/annurev.astro.46.060407.145222}, \href
  {https://ui.adsabs.harvard.edu/abs/2009ARA&A..47..481A} {47, 481}

\bibitem[\protect\citeauthoryear{{Belczynski}, {Kalogera}, {Rasio}, {Taam},
  {Zezas}, {Bulik}, {Maccarone}  \& {Ivanova}}{{Belczynski}
  et~al.}{2008}]{2008ApJS..174..223B}
{Belczynski} K.,  {Kalogera} V.,  {Rasio} F.~A.,  {Taam} R.~E.,  {Zezas} A.,
  {Bulik} T.,  {Maccarone} T.~J.,   {Ivanova} N.,  2008, \mn@doi [\apjs]
  {10.1086/521026}, \href
  {https://ui.adsabs.harvard.edu/abs/2008ApJS..174..223B} {174, 223}

\bibitem[\protect\citeauthoryear{{Belczynski}, {Bulik}, {Fryer}, {Ruiter},
  {Valsecchi}, {Vink}  \& {Hurley}}{{Belczynski}
  et~al.}{2010}]{2010ApJ...714.1217B}
{Belczynski} K.,  {Bulik} T.,  {Fryer} C.~L.,  {Ruiter} A.,  {Valsecchi} F.,
  {Vink} J.~S.,   {Hurley} J.~R.,  2010, \mn@doi [\apj]
  {10.1088/0004-637X/714/2/1217}, \href
  {https://ui.adsabs.harvard.edu/abs/2010ApJ...714.1217B} {714, 1217}

\bibitem[\protect\citeauthoryear{{Belczynski}, {Repetto}, {Holz},
  {O'Shaughnessy}, {Bulik}, {Berti}, {Fryer}  \& {Dominik}}{{Belczynski}
  et~al.}{2016}]{2016ApJ...819..108B}
{Belczynski} K.,  {Repetto} S.,  {Holz} D.~E.,  {O'Shaughnessy} R.,  {Bulik}
  T.,  {Berti} E.,  {Fryer} C.,   {Dominik} M.,  2016, \mn@doi [\apj]
  {10.3847/0004-637X/819/2/108}, \href
  {https://ui.adsabs.harvard.edu/abs/2016ApJ...819..108B} {819, 108}

\bibitem[\protect\citeauthoryear{{Britt}, {Johanson}, {Wood}, {Miller}  \&
  {Michaely}}{{Britt} et~al.}{2021}]{Britt2021}
{Britt} D.,  {Johanson} B.,  {Wood} L.,  {Miller} M.~C.,   {Michaely} E.,
  2021, \mn@doi [\mnras] {10.1093/mnras/stab1570}, \href
  {https://ui.adsabs.harvard.edu/abs/2021MNRAS.505.3844B} {505, 3844}

\bibitem[\protect\citeauthoryear{{Broekgaarden} et~al.,}{{Broekgaarden}
  et~al.}{2019}]{2019MNRAS.490.5228B}
{Broekgaarden} F.~S.,  et~al., 2019, \mn@doi [\mnras] {10.1093/mnras/stz2558},
  \href {https://ui.adsabs.harvard.edu/abs/2019MNRAS.490.5228B} {490, 5228}

\bibitem[\protect\citeauthoryear{{Collins} \& {Sari}}{{Collins} \&
  {Sari}}{2008}]{2008AJ....136.2552C}
{Collins} B.~F.,  {Sari} R.,  2008, \mn@doi [\aj]
  {10.1088/0004-6256/136/6/2552}, \href
  {https://ui.adsabs.harvard.edu/abs/2008AJ....136.2552C} {136, 2552}

\bibitem[\protect\citeauthoryear{{Dominik} et~al.,}{{Dominik}
  et~al.}{2015}]{2015ApJ...806..263D}
{Dominik} M.,  et~al., 2015, \mn@doi [\apj] {10.1088/0004-637X/806/2/263},
  \href {https://ui.adsabs.harvard.edu/abs/2015ApJ...806..263D} {806, 263}

\bibitem[\protect\citeauthoryear{{Ertl}, {Janka}, {Woosley}, {Sukhbold}  \&
  {Ugliano}}{{Ertl} et~al.}{2016}]{2016ApJ...818..124E}
{Ertl} T.,  {Janka} H.~T.,  {Woosley} S.~E.,  {Sukhbold} T.,   {Ugliano} M.,
  2016, \mn@doi [\apj] {10.3847/0004-637X/818/2/124}, \href
  {https://ui.adsabs.harvard.edu/abs/2016ApJ...818..124E} {818, 124}

\bibitem[\protect\citeauthoryear{{Figer}}{{Figer}}{2005}]{2005Natur.434..192F}
{Figer} D.~F.,  2005, \mn@doi [\nat] {10.1038/nature03293}, \href
  {https://ui.adsabs.harvard.edu/abs/2005Natur.434..192F} {434, 192}

\bibitem[\protect\citeauthoryear{{Fragione} \& {Kocsis}}{{Fragione} \&
  {Kocsis}}{2018}]{2018PhRvL.121p1103F}
{Fragione} G.,  {Kocsis} B.,  2018, \mn@doi [\prl]
  {10.1103/PhysRevLett.121.161103}, \href
  {https://ui.adsabs.harvard.edu/abs/2018PhRvL.121p1103F} {121, 161103}

\bibitem[\protect\citeauthoryear{{Fryer}, {Belczynski}, {Wiktorowicz},
  {Dominik}, {Kalogera}  \& {Holz}}{{Fryer} et~al.}{2012}]{2012ApJ...749...91F}
{Fryer} C.~L.,  {Belczynski} K.,  {Wiktorowicz} G.,  {Dominik} M.,  {Kalogera}
  V.,   {Holz} D.~E.,  2012, \mn@doi [\apj] {10.1088/0004-637X/749/1/91}, \href
  {https://ui.adsabs.harvard.edu/abs/2012ApJ...749...91F} {749, 91}

\bibitem[\protect\citeauthoryear{{Giacobbo} \& {Mapelli}}{{Giacobbo} \&
  {Mapelli}}{2018}]{2018MNRAS.480.2011G}
{Giacobbo} N.,  {Mapelli} M.,  2018, \mn@doi [\mnras] {10.1093/mnras/sty1999},
  \href {https://ui.adsabs.harvard.edu/abs/2018MNRAS.480.2011G} {480, 2011}

\bibitem[\protect\citeauthoryear{{Grishin} \& {Perets}}{{Grishin} \&
  {Perets}}{2022}]{gri+22}
{Grishin} E.,  {Perets} H.~B.,  2022, \mn@doi [\mnras] {10.1093/mnras/stac706},
  \href {https://ui.adsabs.harvard.edu/abs/2022MNRAS.512.4993G} {512, 4993}

\bibitem[\protect\citeauthoryear{{Hernquist}}{{Hernquist}}{1990}]{1990ApJ...356..359H}
{Hernquist} L.,  1990, \mn@doi [\apj] {10.1086/168845}, \href
  {https://ui.adsabs.harvard.edu/abs/1990ApJ...356..359H} {356, 359}

\bibitem[\protect\citeauthoryear{{Hoang}, {Naoz}  \& {Kremer}}{{Hoang}
  et~al.}{2020}]{Hoang2020}
{Hoang} B.-M.,  {Naoz} S.,   {Kremer} K.,  2020, \mn@doi [\apj]
  {10.3847/1538-4357/abb66a}, \href
  {https://ui.adsabs.harvard.edu/abs/2020ApJ...903....8H} {903, 8}

\bibitem[\protect\citeauthoryear{{Hobbs}, {Lorimer}, {Lyne}  \&
  {Kramer}}{{Hobbs} et~al.}{2005}]{2005MNRAS.360..974H}
{Hobbs} G.,  {Lorimer} D.~R.,  {Lyne} A.~G.,   {Kramer} M.,  2005, \mn@doi
  [\mnras] {10.1111/j.1365-2966.2005.09087.x}, \href
  {https://ui.adsabs.harvard.edu/abs/2005MNRAS.360..974H} {360, 974}

\bibitem[\protect\citeauthoryear{{Hurley}, {Pols}  \& {Tout}}{{Hurley}
  et~al.}{2000}]{2000MNRAS.315..543H}
{Hurley} J.~R.,  {Pols} O.~R.,   {Tout} C.~A.,  2000, \mn@doi [\mnras]
  {10.1046/j.1365-8711.2000.03426.x}, \href
  {https://ui.adsabs.harvard.edu/abs/2000MNRAS.315..543H} {315, 543}

\bibitem[\protect\citeauthoryear{{Hurley}, {Tout}  \& {Pols}}{{Hurley}
  et~al.}{2002}]{2002MNRAS.329..897H}
{Hurley} J.~R.,  {Tout} C.~A.,   {Pols} O.~R.,  2002, \mn@doi [\mnras]
  {10.1046/j.1365-8711.2002.05038.x}, \href
  {https://ui.adsabs.harvard.edu/abs/2002MNRAS.329..897H} {329, 897}

\bibitem[\protect\citeauthoryear{{Igoshev} \& {Perets}}{{Igoshev} \&
  {Perets}}{2019}]{2019MNRAS.486.4098I}
{Igoshev} A.~P.,  {Perets} H.~B.,  2019, \mn@doi [\mnras]
  {10.1093/mnras/stz1024}, \href
  {https://ui.adsabs.harvard.edu/abs/2019MNRAS.486.4098I} {486, 4098}

\bibitem[\protect\citeauthoryear{{Juri{\'c}} et~al.,}{{Juri{\'c}}
  et~al.}{2008}]{2008ApJ...673..864J}
{Juri{\'c}} M.,  et~al., 2008, \mn@doi [\apj] {10.1086/523619}, \href
  {https://ui.adsabs.harvard.edu/abs/2008ApJ...673..864J} {673, 864}

\bibitem[\protect\citeauthoryear{{Kopparapu}, {Hanna}, {Kalogera},
  {O'Shaughnessy}, {Gonz{\'a}lez}, {Brady}  \& {Fairhurst}}{{Kopparapu}
  et~al.}{2008}]{2008ApJ...675.1459K}
{Kopparapu} R.~K.,  {Hanna} C.,  {Kalogera} V.,  {O'Shaughnessy} R.,
  {Gonz{\'a}lez} G.,  {Brady} P.~R.,   {Fairhurst} S.,  2008, \mn@doi [\apj]
  {10.1086/527348}, \href
  {https://ui.adsabs.harvard.edu/abs/2008ApJ...675.1459K} {675, 1459}

\bibitem[\protect\citeauthoryear{{Kroupa}}{{Kroupa}}{2001}]{2001MNRAS.322..231K}
{Kroupa} P.,  2001, \mn@doi [\mnras] {10.1046/j.1365-8711.2001.04022.x}, \href
  {https://ui.adsabs.harvard.edu/abs/2001MNRAS.322..231K} {322, 231}

\bibitem[\protect\citeauthoryear{{Leigh} et~al.,}{{Leigh}
  et~al.}{2018}]{2018MNRAS.474.5672L}
{Leigh} N.~W.~C.,  et~al., 2018, \mn@doi [\mnras] {10.1093/mnras/stx3134},
  \href {https://ui.adsabs.harvard.edu/abs/2018MNRAS.474.5672L} {474, 5672}

\bibitem[\protect\citeauthoryear{{Madau} \& {Dickinson}}{{Madau} \&
  {Dickinson}}{2014}]{2014ARA&A..52..415M}
{Madau} P.,  {Dickinson} M.,  2014, \mn@doi [\araa]
  {10.1146/annurev-astro-081811-125615}, \href
  {https://ui.adsabs.harvard.edu/abs/2014ARA&A..52..415M} {52, 415}

\bibitem[\protect\citeauthoryear{{McKernan}, {Ford}, {Lyra}  \&
  {Perets}}{{McKernan} et~al.}{2012}]{2012MNRAS.425..460M}
{McKernan} B.,  {Ford} K.~E.~S.,  {Lyra} W.,   {Perets} H.~B.,  2012, \mn@doi
  [\mnras] {10.1111/j.1365-2966.2012.21486.x}, \href
  {https://ui.adsabs.harvard.edu/abs/2012MNRAS.425..460M} {425, 460}

\bibitem[\protect\citeauthoryear{{Michaely}}{{Michaely}}{2021}]{michaely2021a}
{Michaely} E.,  2021, \mn@doi [\mnras] {10.1093/mnras/staa3623}, \href
  {https://ui.adsabs.harvard.edu/abs/2021MNRAS.500.5543M} {500, 5543}

\bibitem[\protect\citeauthoryear{{Michaely} \& {Perets}}{{Michaely} \&
  {Perets}}{2016}]{2016MNRAS.458.4188M}
{Michaely} E.,  {Perets} H.~B.,  2016, \mn@doi [\mnras] {10.1093/mnras/stw368},
  \href {https://ui.adsabs.harvard.edu/abs/2016MNRAS.458.4188M} {458, 4188}

\bibitem[\protect\citeauthoryear{{Michaely} \& {Perets}}{{Michaely} \&
  {Perets}}{2019}]{2019ApJ...887L..36M}
{Michaely} E.,  {Perets} H.~B.,  2019, \mn@doi [\apjl]
  {10.3847/2041-8213/ab5b9b}, \href
  {https://ui.adsabs.harvard.edu/abs/2019ApJ...887L..36M} {887, L36}

\bibitem[\protect\citeauthoryear{{Michaely} \& {Perets}}{{Michaely} \&
  {Perets}}{2020}]{mic+20}
{Michaely} E.,  {Perets} H.~B.,  2020, \mn@doi [\mnras]
  {10.1093/mnras/staa2720}, \href
  {https://ui.adsabs.harvard.edu/abs/2020MNRAS.498.4924M} {498, 4924}

\bibitem[\protect\citeauthoryear{{Michaely} \& {Shara}}{{Michaely} \&
  {Shara}}{2021}]{michaely2021}
{Michaely} E.,  {Shara} M.~M.,  2021, \mn@doi [\mnras] {10.1093/mnras/stab339},
  \href {https://ui.adsabs.harvard.edu/abs/2021MNRAS.502.4540M} {502, 4540}

\bibitem[\protect\citeauthoryear{{Michaely}, {Ginzburg}  \&
  {Perets}}{{Michaely} et~al.}{2016}]{2016arXiv161000593M}
{Michaely} E.,  {Ginzburg} D.,   {Perets} H.~B.,  2016, arXiv e-prints, \href
  {https://ui.adsabs.harvard.edu/abs/2016arXiv161000593M} {p. arXiv:1610.00593}

\bibitem[\protect\citeauthoryear{{Neijssel} et~al.,}{{Neijssel}
  et~al.}{2019}]{2019MNRAS.490.3740N}
{Neijssel} C.~J.,  et~al., 2019, \mn@doi [\mnras] {10.1093/mnras/stz2840},
  \href {https://ui.adsabs.harvard.edu/abs/2019MNRAS.490.3740N} {490, 3740}

\bibitem[\protect\citeauthoryear{{Perets} \& {Kouwenhoven}}{{Perets} \&
  {Kouwenhoven}}{2012}]{2012ApJ...750...83P}
{Perets} H.~B.,  {Kouwenhoven} M.~B.~N.,  2012, \mn@doi [\apj]
  {10.1088/0004-637X/750/1/83}, \href
  {https://ui.adsabs.harvard.edu/abs/2012ApJ...750...83P} {750, 83}

\bibitem[\protect\citeauthoryear{Peters}{Peters}{1964}]{PhysRev.136.B1224}
Peters P.~C.,  1964, \mn@doi [Phys. Rev.] {10.1103/PhysRev.136.B1224}, 136,
  B1224

\bibitem[\protect\citeauthoryear{{Pols}, {Schr{\"o}der}, {Hurley}, {Tout}  \&
  {Eggleton}}{{Pols} et~al.}{1998}]{8135443}
{Pols} O.~R.,  {Schr{\"o}der} K.-P.,  {Hurley} J.~R.,  {Tout} C.~A.,
  {Eggleton} P.~P.,  1998, \mn@doi [\mnras] {10.1046/j.1365-8711.1998.01658.x},
  \href {https://ui.adsabs.harvard.edu/abs/1998MNRAS.298..525P} {298, 525}

\bibitem[\protect\citeauthoryear{{Repetto}, {Davies}  \&
  {Sigurdsson}}{{Repetto} et~al.}{2012}]{2012MNRAS.425.2799R}
{Repetto} S.,  {Davies} M.~B.,   {Sigurdsson} S.,  2012, \mn@doi [\mnras]
  {10.1111/j.1365-2966.2012.21549.x}, \href
  {https://ui.adsabs.harvard.edu/abs/2012MNRAS.425.2799R} {425, 2799}

\bibitem[\protect\citeauthoryear{{Rodriguez}, {Chatterjee}  \&
  {Rasio}}{{Rodriguez} et~al.}{2016}]{2016PhRvD..93h4029R}
{Rodriguez} C.~L.,  {Chatterjee} S.,   {Rasio} F.~A.,  2016, \mn@doi [\prd]
  {10.1103/PhysRevD.93.084029}, \href
  {https://ui.adsabs.harvard.edu/abs/2016PhRvD..93h4029R} {93, 084029}

\bibitem[\protect\citeauthoryear{{Rodriguez}, {Amaro-Seoane}, {Chatterjee},
  {Kremer}, {Rasio}, {Samsing}, {Ye}  \& {Zevin}}{{Rodriguez}
  et~al.}{2018}]{2018PhRvD..98l3005R}
{Rodriguez} C.~L.,  {Amaro-Seoane} P.,  {Chatterjee} S.,  {Kremer} K.,  {Rasio}
  F.~A.,  {Samsing} J.,  {Ye} C.~S.,   {Zevin} M.,  2018, \mn@doi [\prd]
  {10.1103/PhysRevD.98.123005}, \href
  {https://ui.adsabs.harvard.edu/abs/2018PhRvD..98l3005R} {98, 123005}

\bibitem[\protect\citeauthoryear{{Rozner} \& {Perets}}{{Rozner} \&
  {Perets}}{2022}]{2022arXiv220301330R}
{Rozner} M.,  {Perets} H.~B.,  2022, arXiv e-prints, \href
  {https://ui.adsabs.harvard.edu/abs/2022arXiv220301330R} {p. arXiv:2203.01330}

\bibitem[\protect\citeauthoryear{{Samsing}}{{Samsing}}{2018}]{2018PhRvD..97j3014S}
{Samsing} J.,  2018, \mn@doi [\prd] {10.1103/PhysRevD.97.103014}, \href
  {https://ui.adsabs.harvard.edu/abs/2018PhRvD..97j3014S} {97, 103014}

\bibitem[\protect\citeauthoryear{{Samsing}, {MacLeod}  \&
  {Ramirez-Ruiz}}{{Samsing} et~al.}{2014}]{2014ApJ...784...71S}
{Samsing} J.,  {MacLeod} M.,   {Ramirez-Ruiz} E.,  2014, \mn@doi [\apj]
  {10.1088/0004-637X/784/1/71}, \href
  {https://ui.adsabs.harvard.edu/abs/2014ApJ...784...71S} {784, 71}

\bibitem[\protect\citeauthoryear{{Sana} et~al.,}{{Sana}
  et~al.}{2012}]{2012Sci...337..444S}
{Sana} H.,  et~al., 2012, \mn@doi [Science] {10.1126/science.1223344}, \href
  {https://ui.adsabs.harvard.edu/abs/2012Sci...337..444S} {337, 444}

\bibitem[\protect\citeauthoryear{{Stephan}, {Naoz}, {Ghez}, {Witzel},
  {Sitarski}, {Do}  \& {Kocsis}}{{Stephan} et~al.}{2016}]{Stephan2016}
{Stephan} A.~P.,  {Naoz} S.,  {Ghez} A.~M.,  {Witzel} G.,  {Sitarski} B.~N.,
  {Do} T.,   {Kocsis} B.,  2016, \mn@doi [\mnras] {10.1093/mnras/stw1220},
  \href {https://ui.adsabs.harvard.edu/abs/2016MNRAS.460.3494S} {460, 3494}

\bibitem[\protect\citeauthoryear{{Stevenson}, {Vigna-G{\'o}mez}, {Mandel},
  {Barrett}, {Neijssel}, {Perkins}  \& {de Mink}}{{Stevenson}
  et~al.}{2017}]{2017NatCo...814906S}
{Stevenson} S.,  {Vigna-G{\'o}mez} A.,  {Mandel} I.,  {Barrett} J.~W.,
  {Neijssel} C.~J.,  {Perkins} D.,   {de Mink} S.~E.,  2017, \mn@doi [Nature
  Communications] {10.1038/ncomms14906}, \href
  {https://ui.adsabs.harvard.edu/abs/2017NatCo...814906S} {8, 14906}

\bibitem[\protect\citeauthoryear{{Tagawa}, {Saitoh}  \& {Kocsis}}{{Tagawa}
  et~al.}{2018}]{2018PhRvL.120z1101T}
{Tagawa} H.,  {Saitoh} T.~R.,   {Kocsis} B.,  2018, \mn@doi [\prl]
  {10.1103/PhysRevLett.120.261101}, \href
  {https://ui.adsabs.harvard.edu/abs/2018PhRvL.120z1101T} {120, 261101}

\bibitem[\protect\citeauthoryear{{The LIGO Scientific Collaboration}
  et~al.,}{{The LIGO Scientific Collaboration}
  et~al.}{2021a}]{2021arXiv211103606T}
{The LIGO Scientific Collaboration} et~al., 2021a, arXiv e-prints, \href
  {https://ui.adsabs.harvard.edu/abs/2021arXiv211103606T} {p. arXiv:2111.03606}

\bibitem[\protect\citeauthoryear{{The LIGO Scientific Collaboration}
  et~al.,}{{The LIGO Scientific Collaboration}
  et~al.}{2021b}]{2021arXiv211103634T}
{The LIGO Scientific Collaboration} et~al., 2021b, arXiv e-prints, \href
  {https://ui.adsabs.harvard.edu/abs/2021arXiv211103634T} {p. arXiv:2111.03634}

\bibitem[\protect\citeauthoryear{{Vigna-G{\'o}mez} et~al.,}{{Vigna-G{\'o}mez}
  et~al.}{2018}]{2018MNRAS.481.4009V}
{Vigna-G{\'o}mez} A.,  et~al., 2018, \mn@doi [\mnras] {10.1093/mnras/sty2463},
  \href {https://ui.adsabs.harvard.edu/abs/2018MNRAS.481.4009V} {481, 4009}

\bibitem[\protect\citeauthoryear{{Wen}}{{Wen}}{2003}]{2003ApJ...598..419W}
{Wen} L.,  2003, \mn@doi [\apj] {10.1086/378794}, \href
  {https://ui.adsabs.harvard.edu/abs/2003ApJ...598..419W} {598, 419}

\makeatother
\end{thebibliography}

%%%%%%%%%%%%%%%%% APPENDICES %%%%%%%%%%%%%%%%%%%%%

\appendix


\bsp
\label{lastpage}
\end{document}